\documentstyle[preprint,prb,aps]{revtex}
\begin{document}
\title{Addition spectrum, persistent current, and spin polarization 
in coupled quantum dot arrays: coherence, correlation, and disorder}
\draft
\author{R. Kotlyar, 
C. A. Stafford,\footnote{Current and Permanent address:
Fakult\"at f\"ur Physik, Albert-Ludwigs-Universit\"at, D-79104 Freiburg,
Germany.}
 and S. Das Sarma}
\address{Department of Physics,
University of Maryland, College Park, MD 20742-4111, USA}

\date{\today}
\maketitle
\begin{abstract}
The ground state persistent current and electron addition spectrum in 
 two-dimensional  quantum dot arrays 
 and one-dimensional  quantum dot rings,  
pierced by an external
 magnetic flux, are investigated using 
 the extended Hubbard model.
 The collective multidot problem is shown
 to map exactly into the strong field noninteracting 
 finite-size Hofstadter  butterfly problem 
 {\em at the spin polarization transition}. 
The finite size 
Hofstadter problem is discussed, and an analytical solution
 for limiting values of flux is obtained. In weak fields 
we predict novel flux periodic oscillations in the spin 
component along the quantization axis with a periodicity 
given by 
  $\nu\ h/e$ 
($\nu \le 1$). 
 The sensitivity of the calculated persistent
 current to interaction and disorder is shown to reflect the 
intricacies of various Mott-Hubbard quantum phase transitions in two-
dimensional systems: the persistent current is suppressed 
 in the antiferromagnetic Mott-insulating phase governed by
 intradot Coulomb interactions; 
the persistent current is maximized at the spin density  wave - charge density wave transition driven by the nearest neighbor
 interdot interaction; 
 the Mott-insulating phase persistent current is 
enhanced  by the  long-range interdot interactions to its
 noninteracting value; 
the strong suppression of the noninteracting current in the
 presence of random disorder is seen only at large disorder
 strengths; at half filling even a relatively weak intradot Coulomb 
interaction 
 enhances the disordered
 noninteracting system persistent current;
in general, the suppression of the persistent current by disorder is less significant in the presence of the long-range interdot Coulomb interaction.
\end{abstract}
\newpage

\section{Introduction}
In this paper we consider an array of coherently 
coupled\cite{transportexp,exprings,ashoori,pcurexp,lenssen} 
semiconductor quantum dots arranged in finite 
two-dimensional (2D) square lattices or one-dimensional (1D) rings 
pierced by a magnetic flux oriented normal to the lattice or the ring plane.
At low temperatures, these quantum dot arrays may be
 considered 
``artificial molecules'' (with individual quantum dots being
 the ``atomic'' constituents of these artificial molecules) 
because the electron 
 phase coherence length  is
 comparable to the array linear size.
Theoretical work on multidot systems has mostly
 concentrated on the two limiting
 situations: 
coherent dots with no Coulomb
 interaction\cite{kirczenow} and Coulomb blockade of individual 
dots\cite{bookset,middleton}. In this paper, 
we consider quantum dot arrays taking into 
account 
quantum fluctuations arising from interdot hopping, electron-electron 
interaction, and random disorder effects 
through an extended Hubbard-type 
Hamiltonian\cite{dotshubtheory1,dotshubtheory2,kotlyarsdsPCUR}

In a previous paper,\cite{kotlyarsdsPCUR} we reported on our prediction of an {\it{equilibrium persistent current}} in 
finite 2D dot arrays (without any periodic boundary conditions) in the presence of an applied magnetic field transverse to the 2D plane. In this paper 
we provide details and expand on our 
previous work, and present results for the electron addition spectrum and the persistent current in 2D square lattices and 1D rings including 
effects of collective physics arising from the multiple dot structure of the system within a simple model for the
 single-particle physics\cite{fock} of the individual quantum dots. 
One of our primary motivations is to understand {\em lattice} 
effects on the persistent current and the electron addition spectrum, 
the lattice here being the {\em artificial lattice} defining the 2D or the 
1D quantum dot array (with the typical lattice constant in the $20-200$ nm 
range). One goal is to identify experimentally observable features of various Mott-Hubbard 
quantum phase transitions (including realistic disorder and interaction effects) in semiconductor quantum dot arrays.

The importance of electron-electron interaction  has been  
stressed\cite{imryrev,cheung,shastrysutherland,pcurtheory1Dfor,pcurtheory1Dagainst,buttikerstafford} in the literature in the context of 
 persistent current  experiments in 1D gold and semiconductor rings\cite{pcurexp}. The magnitude of the persistent current\cite{pcurexp} in 
 {\em disordered} gold rings was found to be one to two orders of magnitude larger than 
that theoretically predicted, 
whereas in {\em clean} semiconductor rings the magnitude of the 
persistent current 
was found to be in a good agreement with the theoretically 
predicted  simple noninteracting value of $e\ v_{F}/L$ 
(with $v_{F}$ being the Fermi velocity of electrons moving in a 
ring of length $L$, in our notation $L$ denotes the size of the system as 
defined by the total number of dots in it). Although 
the interplay between disorder and Coulomb interaction 
in determining the magnitude of the persistent current in ring topologies is 
the subject of many recent theoretical investigations\cite{pcurtheory1Dfor,pcurtheory1Dagainst,thickrings}, the issue
 remains unsettled. 
%A basic theoretical controversy is that the 
%perturbative theories claim that an enhancement of the 
%magnitude of the disordered persistent current in the 
%presence of the short-range Coulomb interactions 
%can be sufficient to account for the 
%experimental value\cite{pcurtheory1Dfor}, whereas other 
%theoretical investigations 
%based on studies of the 
%spinless fermion systems disagree 
%with this\cite{pcurtheory1Dagainst}. 
%An additional mechanism for an enhancement of the
% magnitude of the persistent current is to consider thick
% rings with many correlated channels\cite{thickrings}.
Disorder and interaction effects are naturally included in our 
Mott-Hubbard model of finite quantum dot arrays, and 
we will comment on their influence on the persistent current.

The paper is organized as follows. In Sec. II we define  
and describe the extended Mott-Hubbard Hamiltonian\cite{dotshubtheory1} 
which forms the basis of our theoretical description of 
the collective physics in 
 finite quantum dot lattices.  In Sec. III 
we present our calculated electron addition spectrum as a function of the
externally applied magnetic flux for a $3 \times 3$ lattice and 
an $L=9-$site ring, and identify the main features of these results 
which are
 studied in the subsequent sections. In Sec. IV we clarify the physical
 meaning of the 
magnetic field dependence of the addition spectrum by 
demonstrating the equivalence between the derivative of the Hamiltonian 
with respect to the flux and the magnetization 
density (or equivalently the persistent current) operators. In Sec. V we study the energy spectrum and the persistent current of the finite open 2D $\sqrt{L} \times \sqrt{L}$ tight-binding lattices of $L$ noninteracting quantum dots. We also discuss in this context the Hofstadter spectrum
 of an infinite tight-binding lattice. 
We establish a connection 
between the ground state magnetization in the lattice and in the continuum 2D system by identifying the different regions in the energy 
spectrum in a lattice, and illustrate them with the calculated 
distributions of the persistent current on a $15 \times 15$ lattice. We also solve the problem exactly for lattices of arbitrary sizes in the two limiting situations of the flux $\phi=0$ and $\phi=0.5\ \phi_{0}$ per unit
 cell (with $\phi_{0}=h/e$ being the fundamental flux unit),
  and classify the $\phi=0$ states using 
 group theory and perturbation theory.
We also discuss in Sec. V our results for the noninteracting persistent current in 1D rings. 
In Sec. VI we study the electron addition spectra of interacting lattices in a magnetic field. In Sec. VIA  we 
study intradot Coulomb interaction effects 
 within the minimal Hubbard model approximation.
 In particular, we study the {\it{periodic}} oscillations of the
 component $S_{z}$ of the total ground state electron spin along the quantization $(z)$ axis of the 2D Hubbard model using the Lanczos exact
 diagonalization technique, and by solving 
 the Bethe ansatz equations in 1D rings. We reanalyze the 
 persistent current results in a 
1D Hubbard ring, and find that the interacting system behaves as a 
{\em single particle} by changing its total orbital momentum sequentially as a function of the flux. The details of our Bethe ansatz 
analysis for the 1D Hubbard ring 
spectrum are given in Appendix A. 
In Sec. VIA we also discuss the finite size realization of the 
Mott-Hubbard metal-insulator transition and magnetic ordering
 in finite $2 \times 2$, $3 \times 2$, $4 \times 2$ and $3 \times 3$ quantum dot clusters (with $L=4,\ 6,\ 8,\ 9$ dots respectively in the system). 
In Sec. VIB we discuss the spin density wave-charge density wave ordering transition in a half-filled 2D $3 \times 3$ array in the presence of the nearest neighbor interactions. We find an enhancement of the
persistent current along the transition line. In Sec. VII, we consider 
 random disorder effects on the 2D persistent current. We find that the noninteracting persistent current
 as a function of the disorder strength shows a behavior similar to that of the conductivity\cite{lee}: it is strongly suppressed only at large disorder strengths. We obtain an empirical 
scaling function for the persistent current as a function of the disorder strength. Finally, we discuss the effects of having both disorder and 
Coulomb interaction on the 2D persistent current in the $3\times3$ array 
by using the Lanczos exact
 diagonalization technique. We conclude with a summary of our results
in Sec. VIII.

\section{Model}
We model an {\it{isolated}} finite system (``array'') of coherently coupled 
(nominally identical) semiconductor quantum dots arranged in one-dimensional rings or two-dimensional rectangular (``square'') lattices at zero 
temperature.
We assume that the charging of an otherwise electrically neutral quantum dot array with a fixed number $N$ of excess quasiparticles is accomplished through the tunneling of the quasiparticles from a nearby backgate electrode. 
The excess charges are shared among all dots in the array in a molecular-like fashion by 
quantum-mechanical tunneling, i.e., our quantum dot array is coherent.
The equilibrium properties of quantum dot arrays  we investigate in this work can in principle be measured through experiments on 
 semiconductor dot arrays using available experimental techniques: tunneling transport spectroscopy\cite{transportexp},
 equilibrium capacitance 
spectroscopy\cite{ashoori}, and equilibrium magnetization measurements\cite{pcurexp}.

The Hamiltonian of an isolated array of $L$ coupled 
quantum dots is given by the sum of three terms: 
\begin{equation}
\label{allham}
H_{array}=(KE)_{sp}+(KE)_{hop}+V_{int}.
\end{equation} 

The $(KE)_{sp}$ term in Eq. \ref{allham} 
includes all intradot single-particle effects (including confinement 
 contribution) and is 
 the 
sum of the 
single-particle Hamiltonians of the 
individual quantum dots of the array. We model  GaAs 
quantum dots at zero temperature. The characteristic 
size of each dot is taken to be $D \gg a_{B}$ 
where $a_{B}(\approx  100\ \AA)$ is the effective Bohr radius of 
the bulk GaAs material. All effects of the electron-electron 
interactions {\em within a neutral dot} are absorbed in the effective values of 
quasiparticle parameters giving
 the effective mass $m^{\star}=0.067 \ m_{e}$ 
and the effective $g$-factor $g^{*}= 0.2 \ g_{e}$ in a standard band 
description for the single-particle 
energy levels within the dot. The confinement potential 
of a quantum  dot  is known to be approximately 
parabolic\cite{kumar}. 
Due to quantum confinement in the dot, a continuous
 conduction band for the  excess quasiparticles is a discrete 
series of single-particle energy levels 
$\varepsilon_{\alpha}$ where $\alpha$ denotes a 
single-particle state including spin. The  
single-particle intradot level spacing at zero field is taken to 
be $\Delta = \hbar \omega_{0}$ (with $\omega_{0}$ essentially being a 
harmonic oscillator frequency). 
In our work we consider only the lowest intradot states near the Fermi 
level. 
In the occupation basis 
of states $| \alpha\rangle$ the $(KE)_{sp}$ term is 
written in the second quantized notation as 
\begin{equation}
\label{keintradot}
(KE)_{sp}=\sum_{i,\alpha}^{L} \varepsilon_{i\alpha}(B) 
 {{c}^{\dag}}_{i\alpha} c_{i\alpha}.
\end{equation}
In Eq. \ref{keintradot} the summation is over all dots $i$; ${{c}^{\dag}}_{i\alpha} (c_{i\alpha})$ is a creation (annihilation) operator for a quasiparticle on the $i$th dot
 in a state $\alpha$.  The single-particle magnetic 
field dependence is included 
in Eq. \ref{keintradot} through the usual Fock-Darwin-Zeeman scheme as
\begin{equation}
\label{spenergy}
\varepsilon_{i\alpha} \equiv
\varepsilon_{\alpha}=\hbar [(\omega_{c}/2)^{2}+{\omega_{0}}^{2}]^{1/2}
+ (-1)^{\alpha} g_{e} \mu_{B}B/2.
\end{equation}
with $\omega_{c}=eB/m^{*}$ being the cyclotron frequency. 
We consider a 
single spin-split level per dot, and set  $\alpha=1,2$ in Eq. \ref{spenergy} to 
correspond to the spin up/down lowest confined quantum dot level. 
We neglect  correlations 
arising from single-particle level crossings that have to be taken into account 
for fields larger 
than $B=[g^{\star} (g^{\star}+2 m/m^{*})]^{-1/2}(\Delta/\mu_{B})$, and 
concentrate on the collective physics in the array. 

The quasiparticles are allowed to tunnel (``hop'') between the same 
single-particle states $|\alpha \rangle$ in the dots with the 
tunneling amplitudes $t_{\alpha} \equiv t$. In the tight-binding approximation, 
keeping only nearest neighbor tunneling, the tunneling energy is given by 
\begin{equation}
(KE)_{hop}=\sum_{<i,j>, \alpha} 
(t_{\alpha}e^{i\phi_{ij}} {{c}^{\dag}}_{i\alpha} 
c_{j\alpha} + h.c.), \\
\label{keinterdot}
\end{equation}
where
$\phi_{ij}= \frac{e}{\hbar } \int_{ij} \vec{A} \cdot \vec{l_{ij}}$ is 
the Peierls phase factor\cite{peierls}, 
 with  $\vec{A}$ as the magnetic vector potential. 
The indices $i,\ j$ denote the spatial positions of the dots.
The $(KE)_{hop}$ term defines the topology of the array. We model a one-dimensional ring of L quantum dots with a total magnetic flux $\phi$ piercing its 
enclosed area and a two-dimensional square lattice of $L=L_{x} \times L_{y}$ with open boundary conditions with a magnetic flux $\phi$ piercing {\em each unit cell} of the lattice. 

The third term in Eq. (\ref{allham}) defines the intra- and interdot 
Coulomb interactions between quasiparticles 
\begin{equation}
\label{interactions} 
V_{int}=
\sum_{ij} \frac{V_{ij}}{2} {\hat{\rho}}_{i} {\hat{\rho}}_{j}, 
\end{equation}
where ${\hat{\rho}}_{i}=\sum_{\alpha}{{c}^{\dag}}_{i\alpha} c_{i\alpha}$ is the number operator for the $i$th quantum dot.   The interaction constants 
are related to the capacitance matrix of the quantum dot array by
$V_{ij}= (C^{-1})_{ij}$, where
$C_{ii}=C_{g}+N_{i}  C$ and $C_{ij}=-C$ for nearest neighbor dots.
\cite{dotshubtheory1,bookset,middleton}
The capacitance $C_{g}$ represents the capacitance of a quantum dot 
with respect to the
external gates, while $C$ denotes the capacitive coupling
between the $i$th dot and  
the $N_{i}$ neighboring dots.  Eq.\ (\ref{allham}) thus has the form of
an extended Hubbard model with screened long-range interactions.  For
$C \ll C_g$, the interaction matrix elements fall off as $V_{ij} \sim
U (C/C_g)^{|i-j|}$.
We include  
the effects of short range interactions, keeping only the on-site interaction
$U$ and the nearest neighbor interaction $V$. 
We use $ t = 0.1$ meV, 
$\Delta= 3t$, and $U=10t$ in our calculations (unless otherwise stated) as 
a representative set of Hubbard parameters describing the GaAs dot arrays.
 The Hamiltonian given by Eqs. \ref{allham} - \ref{interactions} 
has  been used earlier to  describe coherent 1D 
quantum dot chains with open or periodic boundary 
conditions\cite{dotshubtheory1,dotshubtheory2}. 
(Inclusion of disorder in our model will be discussed in Sec. VII.)
We calculate the electron addition spectrum by doing an
exact diagonalization of Eq. \ref{allham} in the subspace of the total 
number of quasiparticles $N$ in the array and the total 
spin component $S_{z}(=- \frac{1}{2} (N-M)+\frac{1}{2} M$ 
with $M$ being a number of spin-up electrons) 
along the external magnetic  
 field $\vec{B}$. The 
Hilbert space of Eq. \ref{allham} with fixed $N$ and $S_{z}$ 
grows exponentially with the 
system size. We use the Lanczos 
 method\cite{cullum} for our exact 
diagonalization of Eq. \ref{allham}, and 
carry out a ground state energy minimization over $S_{z}$ to find the
stable ground state for a given $N$. The largest matrix size 
 that we have considered is $15876^{2}$ which corresponds to 
the interacting $L=9$ (or $3 \times 3$) dot array at half-filling for $N=L=9$. We also 
calculate the energy spectrum of 
one-dimensional rings in the limit $C=0$
($V_{ii}=U\neq 0,\ V_{ij}=0$) by
solving numerically the corresponding Bethe ansatz exact solution equations. 

\section{Electron addition spectrum}
By definition the chemical potential $\mu_{N}$ of the 
array is given by 
\begin{equation}
\label{chempotdef}
\mu_{N}=E_{0}(N)-E_{0}(N-1),
\end{equation}
where $E_{0}(N)$ is the minimum eigenvalue (i.e. the ground state 
energy) of Eq. \ref{allham} 
in a space of fixed $N$ and all allowed $M$. 
The addition spectrum of the system is the $\mu_{N} -N$ plot, 
which we show in Figs. 1 and 2.
The calculated chemical potential 
of a $3 \times 3$ array and a $9-$site ring in the minimal 
Hubbard model approximation ($V_{ii}=U\neq 0,\ V_{ij}=0$) 
is shown  as a function of the applied flux $\phi$
in Fig. 1.
Each curve in Fig. 1a (c) traces the chemical potential 
$\mu_{N}$ of a $3 \times 3$ array ($L=9$ sites ring) with $N$ 
electrons as a function of the magnetic flux $\phi / \phi_{0}$  
through a unit cell (the ring), with $\phi_{0}=h/e$. The $z$-component 
($S_{z}$) of the total spin of the 
corresponding ground state of the system is shown in Fig. 1 (b) 
and (d), and the 
critical magnetic flux for full spin polarization in the array and the ring is 
given as an inset in Fig. 1 (a) and (c), respectively.

The three main features of the results shown in 
Fig. 1 are: (a) the chemical potential spectrum evolves  
 with the maximization of the total spin polarization 
in the system; (b) apart from an aperiodic 
single-particle background contribution, $\mu(N)$ is a periodic function  
of the magnetic flux with a flux periodicity of $ \gamma\ \phi_{0}$ with  
 $\gamma \le 1$; 
(c) the spin polarization transition of the system occurs 
through the cycles of the {\it{periodic}} $S_{z}$ oscillations in the
 weak magnetic field region.

The result (a) is a trivial outcome of the 
 minimal Hubbard model: a dominant 
Zeeman energy term $g^{*}\mu B$ leads to single-occupancy, effectively 
suppressing the 
presence of the  Hubbard $U$ term  in Eq. \ref{allham}. 
The chemical potential spectrum therefore behaves as that of noninteracting 
spinless fermions in the maximum spin-polarization region. 
The energy spectra of the $9-$site systems 
 plotted in Fig. 2(a) and in Fig. 7 for noninteracting 
quasiparticles [setting 
$\varepsilon_{\downarrow}=\Delta$, $N=M$ in Eq. \ref{allham}] can be 
directly compared with the strong field regions in Figs. 1(a) and (c). 
 In the weak field region the single-particle 
physics can again be distinguished from the collective physics, and the 
spectrum in this regime is plotted in Figs. 2(b) and (c) for the array and 
the ring, respectively. In Figs. 2(b) and (c) 
we set 
$\varepsilon_{\downarrow}=\Delta, \ \varepsilon_{\uparrow}=\Delta+\delta$, 
where $\delta/\Delta (=0.03) \ll 1$, and $M$ 
corresponds to the minimum eigenvalue of Eq. \ref{allham} 
for a given $N$. (We keep a small field-independent shift 
$\delta$ between opposite spin single-particle states for our 
later discussion of the $S_{z}$ oscillations.) The parameters of the Hamiltonian that we use in 
Eqs. (\ref{allham})-(\ref{interactions}) determine the value of the critical magnetic flux $\phi_{c}$ needed to produce full spin polarization in the  
$3 \times 3$ array to be 
$\phi_{c} \approx 8 t^{2} a^{2} /(g^{*} \mu_{B}U) \approx 52 \phi_{0}$ with a lattice constant of 
 $a=280$ nm. For illustrative purposes we rescaled the magnetic 
flux by $1/32$ to show the full behavior of the spectrum on 
a single scale in Figs. 1(a) - (d). (This rescaling physically corresponds to 
rescaling the lattice constant to $\approx a/6$.)

\section{Persistent current}
We first clarify the physical meaning of the intricate
magnetic field dependence of the energy 
spectrum shown in Figs. 1 and 2. It is 
well-known\cite{byers} that gauge invariance
along with the single-valuedness of the electron wavefunction allows 
for the existence
of a ground state persistent current in normal metal rings threading an
external magnetic flux. The intrinsic magnetic moment associated with
this persistent current, which is proportional to the persistent current 
itself in
1D rings, is an oscillatory function of the external flux 
with a period equal to the elementary flux quantum $\phi_{0}$. 
The existence of such an oscillatory persistent current in
normal metal rings has been
experimentally verified\cite{pcurexp}.

In finite 
 2D
{\em continuous} systems no magnetization is expected 
classically.\cite{peierlssurpr} It is argued that the magnetization 
due to electron orbits along the edge of the sample exactly cancels the
 magnetization 
arising from the bulk orbits. In a quantum-mechanical description, 
 however, 
the contribution from the edge states is 
expected to be statistically insignificant\cite{peierlssurpr}, and 
the bulk contributions lead to the famous Landau diamagnetism in 
macroscopic 2D systems. 
In mesoscopic systems the
 phase coherence length $L_{\phi}$ is comparable 
to the linear system size $L$, and as was shown in several 
theoretical papers\cite{altshuler,beenakkerAB}, the edge states in this situation can carry a persistent current creating a paramagnetic moment 
in the continuous 2D geometries (e.g., a 2D disk-shaped quantum dot) 
comparable in magnitude to the Landau 
diamagnetic term. The Aharonov-Bohm effect leads to a flux
 periodicity of the magnetization carried by the edge states effectively forming a 1D ring geometry in continuous 2D mesoscopic systems. 
We characterize the ground state magnetization of finite 2D quantum dot
 arrays 
without resorting to an artificial separation of bulk and edge states 
(which are not really meaningfully distinguishable in small structures) by 
considering  
the lattice model. The lattice spectrum 
 contains the ``edge'' states, the ``bulk'' states, and all other
 electron states given by the superposition of all topologically 
closed electron
 paths in the finite 2D lattice.

The part of the Hamiltonian in Eq. \ref{allham} leading to the 
 persistent current is the kinetic energy ``hopping'' 
$(KE)_{hop}$
 term in Eq. \ref{keinterdot}. The total current operator\cite{mahan} 
is given by the commutator of the Hamiltonian with 
the polarization 
operator 
\begin{equation}
\label{totalcurrent}
{\bf J}_{T} = \frac{\imath}{\hbar} \left[ H, {\bf P} \right] , \ \ 
{\textstyle where} \ \ {\bf P} = - e \sum_{i} {\bf R}_{i} {\hat{n}}_{i}.
\end{equation}
In a discrete lattice subjected to a magnetic field, the above formula becomes
\begin{equation}
\label{currentatflux}
{\bf J}_{T} =  -  \frac{4 \pi}{\phi_{0}} 
\sum_{< i j >, \ \alpha} ({\bf R}_{j}  - {\bf R}_{i})  
 \ {\textstyle Im} \left\{ t_{ij,\alpha}\ e^{\imath \phi_{ij}} 
{{c}^{\dag}}_{i \alpha} c_{j \alpha} \right\}.
\end{equation}
The expectation value of the total current operator is zero in the ground 
state of an isolated finite quantum dot lattice. 
Then each 
divergenceless term in the sum of Eq. \ref{currentatflux} can be 
identified as the current between  two nearest neighbor 
lattice sites $i$ and $j$: 
\begin{equation}
\label{currentij}
{\bf J}_{ij} = - \frac{4 \pi}{\phi_{0}} 
\ \sum_{ \ \alpha} \  ({\bf R}_{j}  - {\bf R}_{i})  
\ {\textstyle Im} \  \left\{ t_{ij,\alpha}\ e^{i\phi_{ij}} {{c}^{\dag}}_{i \alpha} c_{j \alpha} \right\}.
\end{equation}
The magnetic moment operator is
\begin{equation}
\label{magmoment}
{\bf M} = \frac{1}{2}   \ 
\int \ {\bf R} \times {\bf J}({\bf R})  \ d^{3} R.
\end{equation}
The magnetization density
operator has 
a non-vanishing $z$-component 
(i.e. along the magnetic field direction) given by
\begin{equation}
\label{magdensity}
m_{z} = \frac{M_{z}}{Area}=
\frac{1}{2\ n_{cells}} \sum_{< i j >}  \ 
J_{ij} \left[ x_{i} (y_{j} - y_{i}) \ - \ y_{i} (x_{j} - x_{i}) \right].
\end{equation}

Apart from the field dependence of single-particle levels in a single dot, 
the derivative of the Hamiltonian in Eq. \ref{allham} with respect to 
the flux is 
\begin{equation}
\label{dhamdflux}
\frac{\partial H}{\partial \Phi} = 
\frac{\partial H}{n_{cells} \partial \phi} = 
\frac{2}{n_{cells}} \sum_{< i j >, \ \alpha} 
\left(\frac{\partial \phi_{ij} }{\partial \phi} \right)
\ {\textstyle Im} \ 
 \left\{t_{ij, \alpha}\ e^{\imath \phi_{ij}} {{c}^{\dag}}_{i \alpha} c_{j \alpha} \right\}.
\end{equation}
In a symmetric gauge, $\vec{A} = (- B \frac{y}{2}, B \frac{x}{2},0)$  in a 
uniform field, and one gets for the Aharonov-Bohm phase $\phi_{ij}$: 
\begin{equation}
\label{dphasedflux}
\frac{\partial \phi_{ij} }{\partial \phi} = 
 \frac{2 \pi}{\phi_{0}} \left[ \frac{x_{i}}{2} (y_{j} - y_{i}) \ - 
\frac{ y_{i}}{2}  (x_{j} - x_{i}) \right].
\end{equation}
The two operators in Eqs. \ref{magdensity} and \ref{dhamdflux}
 are equivalent, and therefore   
\begin{equation}
\label{equiv}
m_{z} \equiv - \frac{\partial H}{n_{cells}\partial \phi}.
\end{equation}
Eq. \ref{equiv} is the
 usual {\em thermodynamic} expression for the magnetization, 
which we have derived here for our microscopic model.

In this paper we use the convention of calling the magnetization density $m_{z}$ the persistent 
current $I$ in both 1D and 2D systems:
\begin{equation}
\label{pcurdef}
I \ \equiv \ m_{z} \ = \ - \frac{\partial H}{n_{cells} \partial \phi}.
\end{equation}
Experimentally, the equilibrium persistent current is usually 
observed by measuring the ground state magnetization.
\section{The non-interacting spectra}
We consider first the Hofstadter problem\cite{hofstadter} of a 
single particle in a magnetic field on an infinite tight-binding lattice. 
In the Landau gauge, $\vec{A}=(0,Bx,0)$, the discrete 
Schr$\ddot{o}$dinger equation in the occupation 
basis $|\alpha \rangle= {\displaystyle{\sum_{(x,y)}}} \psi(x,y)$ is
\begin{equation}
\label{discrSchr}
\psi(x+1,y)+\psi(x-1,y)+
e^{-i 2 \pi \frac{\phi}{\phi_{0}} x} \psi(x,y+1)+
e^{i 2 \pi \frac{\phi}{\phi_{0}} x} \psi(x,y-1)=-E/t \ \psi(x,y),
\end{equation} 
where $\phi$ is the flux through a unit cell, and the lattice 
constant $a$ is taken as the unit length. At $\phi=0$, an infinite system 
described by Eq. \ref{discrSchr} is translationally invariant. 
The coefficients of Eq. \ref{discrSchr} involve only $x$. The $y-$ motion
 separates out\cite{hofstadter} assuming that the $y-$part of the 
wavefunction preserves its $\phi=0$ plane wave form:
\begin{equation}
\label{hofstwfn}
\psi(x,y)=A(k_{y}) \ \psi(k_{y},x) \ e^{\imath k_{y} y}.
\end{equation}
The function $\psi (k_{y}, x)$ is a solution of 
\begin{equation}
\label{harpereq}
\psi(x+1) + \psi(x-1)+ 2\cos \left( 2 \pi \frac{\phi}{\phi_{0}} x- 
k_y \right) \psi(x) = - E/t \ \psi(x).
\end{equation}
Eq. \ref{harpereq} is the well-known Harper equation, 
describing a particle in a 
one-dimensional quasi-periodic potential. An infinite 
tight-binding lattice in a nonzero flux is no longer 
invariant under lattice translations.
It has, however, been shown that it is invariant under magnetic 
translations\cite{Zak} by $q a$, whenever the flux through a 
unit cell is a rational fraction $\phi = \frac{p}{q}\  \phi_{0}$ of 
the fundamental flux quantum with $p$ and $q$ being any 
two integers. 
The Harper equation spectrum has $q$ energy bands at the 
rational fractional values of flux $\phi= \frac{p}{q}\  \phi_{0}$. The magnetic translations by $q a$ define a magnetic unit cell with the 
 total flux 
$\phi=p \phi_{0}$ through the magnetic unit cell. 
At incommensurate flux values (i.e. when $\phi/\phi_{0}$ is not rational)
 the spectrum 
becomes a Cantor set.\cite{hofstadter}
The Cantor set spectrum 
has an infinite number of energy bands that exhibit a self-similar multifractal behavior. The energy spectrum of Eq. \ref{harpereq} is always 
a continuous function of the magnetic flux,\cite{hofstadter} independent of whether $\phi/\phi_{0}$
 is rational or irrational. This enables one to make a direct connection 
between  the
lattice and  the continuum spectra, for example, in identifying Landau bands in the lattice spectrum. 

The basic features of the spectrum\cite{hofstadter} of Eq. \ref{harpereq} 
(which we will refer to as the Hofstadter spectrum), 
which are also pertinent for a finite lattice, are: 
(a) it is periodic in $\phi_{0}$, $E(\phi)=E(\phi + n  \phi_{0})$; 
(b) it is an even function of flux, $E(\phi) = E(-\phi)$; 
(c) both $E(\phi)$ and $-E(\phi)$ belong to the spectrum; 
(d) the spectrum is bounded, $-4t \le E(\phi) \le 4t$. 
A quarter of the Hofstadter
spectrum plotted for rational values of the flux is shown in Fig. 3 where we explicitly label the first two Landau bands. The Landau bands are formed in 
the continuum limit 
when the magnetic length far exceeds the lattice constant, i.e.
 $l_{0}=\sqrt{(1/ 2 \pi) (\phi_{0}/\phi)} \gg 1$. In the tight-binding model, the 
effective mass is $m \approx \hbar^{2}/2t$. The energy levels 
in a lattice for $l_{0} \gg 1$ are approximately given by the
 continuous system Landau level
expression, $E_{n}=\hbar \omega_{c} (n+\frac{1}{2})$ 
with $\hbar \omega_{c}=4 \pi t \ \phi/ \phi_{0}$\cite{sivan}. For 
example, the ratios of the slopes of the 
first three Landau bands in Fig. 3 obey $23.4 : 15.1 : 5.5 \approx 5 : 3 : 1$.
The analytical solution of Eq. \ref{harpereq} for 
rational flux values were recently obtained using the 
Bethe ansatz method.\cite{solutionhofst} 
The complete spectrum is extremely complex, but a general feature of the spectrum which can be seen in Fig. 3 and to which we will 
later return in our study of the 
persistent current in finite systems, is the presence of 
energy bands separated by large gaps.

A quarter of the spectrum for a finite 
$\sqrt{L} \times \sqrt{L} = \ 15 \times 15$ ($L=225$) lattice is shown in Fig. 4. A qualitative similarity between 
the spectra plotted in Figures 3 and 4 was pointed out in the 
literature\cite{sivan}: the presence of similar energy bands in the spectrum, where 
the gaps between the bands are 
filled by the edge states which necessarily exist in finite systems. A finite 
spectrum was studied earlier in connection with the 
Quantum Hall effect and mesoscopic Aharonov-Bohm fluctuations.\cite{sivan} 
The emphasis of these earlier studies was on the part of 
the spectrum where both the Landau 
bands and the edge states can be clearly identified 
(the region from (d) to (e) in Fig. 4). Using Eqs. (\ref{keinterdot}) and 
(\ref{currentij}), we calculate the persistent current distributions in 
finite lattices for {\em all} eigenstates of 
Eq. \ref{discrSchr}. We follow Ref.\onlinecite{sivan} in the identification of different regions of the spectrum and illustrate each region with a sample current and a charge density distribution shown in Fig. 5.
The lower-half band states in the weak field\cite{sivan} regime 
($l_{0} > \sqrt{L}$) labeled (a), (b) and (c) in Figures 4 and 5
 can be considered to be
extended bulk states because the 
paths of the persistent current carried by these states extend 
across the sample with  finite weights  
 both at the boundaries and in the bulk. The sign of the current 
carried by these states oscillates, but as $\phi \rightarrow 0$ 
 it is determined by the degree of the 
degeneracy $g$ of the spectrum at $\phi=0$.
We identify the first three Landau bands in the spectrum 
labeled (d), (e), and (f) in Figures 4 and 5 
(the traces of the 4th and 5th bands can also be seen in the plot). 
The ratios of the slopes of the first three bands are 
comparable to those of the infinite 
tight-binding lattice, $20.2 : 14.1 : 5.3$.
 It is interesting to note that the radii of the 
``large weight'' 
persistent current orbits in Fig. 5 approximately satisfy the 
semiclassical expression for the cyclotron radius of the 
$n$th Landau band, $R_{n}=l_{0}\sqrt{2 n +1}$. The ratios of the 
radii of the persistent current orbits of the first three 
Landau bands in Fig. 5 obey $5: 4 : 2 \approx \sqrt{5}: \sqrt{3}: 1$. 
Thus the continuous system result approximately holds for a finite 
lattice as well: each $n$th Landau orbit accommodates one more flux 
quantum than the $(n-1)$st band orbit.

For larger flux $\phi \ge \frac{\phi_{0}}{2 \pi}$, the 
Landau levels form a more complicated but less 
degenerate pattern. We label the representative 
states (g), (h), and (i) for this region. The distribution of the persistent current carried by these states appears to consist of disconnected orbits, which can be found anywhere in the lattice. Intuitively, the existence of these
disconnected persistent current orbits is consistent
with our general expectation that smaller quantization orbits should not be strongly affected by the confining potential. 

The gaps of the infinite system in Fig. 3 are 
filled with  edge states between the bulk Landau levels and the 
branched Landau levels in Fig. 4. The current distributions of 
the representative edge states 
labeled (j), (k), and (l) in Fig. 4 are shown in Fig. 5.
These edge states have the largest weights concentrated 
near the boundaries in agreement with these 
states being called the ``edge'' states. The edge states in the 
regions between branched Landau levels (state (l) in Figures 4 and 5) 
seem to have a more complicated spatial distribution than the state labeled (j). The persistent current orbits of the state (l) are still extended across the
 lattice. 
The sign of the current carried by edge states of the type (j) is paramagnetic in accordance with the semiclassical argument
given by Peierls\cite{peierlssurpr}.
The flux and energy separation between  these states correspond 
to one added flux quantum through the total area of the sample. This is the origin of the periodic Aharonov-Bohm oscillations superimposed on a regular pattern of de Haas-van Alphen 
oscillations in the magnetization and in the magnetoconductance discussed 
earlier by Sivan and Imry\cite{sivan}.

The total persistent current $I$ of a system of $N$ noninteracting spinless particles for two values of $N$ selected in Figures 4 and 5 is shown in Fig. 6. 
From the symmetries of the 
$L=L^{1/2} \times L^{1/2}$ lattice spectrum, 
it follows that $I(-\phi)=-I(\phi)$, $I(\phi+ n \phi_{0})=I(\phi)$, 
and $I(N)=I(L-N)$. The total current is given by the sum of 
currents carried by all occupied single-particle states, and it can be 
quite different from the persistent current carried at the Fermi energy. 
For example, in Fig. 6(a), at flux $\phi=0.25\ \phi_{0}$, the state at 
the Fermi energy is a 2nd Landau level bulk state (denoted (e) in 
Figures 4 and 5) which carries a diamagnetic 
persistent current. But the total persistent current is 
positive at this value of flux in Fig. 6(a), corresponding to the 
positive contributions from each filled edge state. 
The change in the sign of the current from 
paramagnetic to diamagnetic in the region from 
$\phi \approx 0.1\ \phi_{0}$ to $\phi \approx 0.2\ \phi_{0}$ 
corresponds to merging all first 26 levels into the lowest 
Landau band. This merging of edge states into bulk states was discussed in 
Ref.\onlinecite{sivan}. 
A similar change of the sign of the persistent current can be seen in Fig. 6(b) for the region of flux from $\phi \approx 0.3\ \phi_{0}$ 
to $\phi \approx 0.37\ \phi_{0}$, where the 
Fermi Energy crosses from the branched Landau states through an edge state into the lower Landau level branched states.

As  mentioned above, an analytical solution for the eigenvalues of the Hofstadter problem (Eq. \ref{harpereq}) at rational values of flux on an infinite tight-binding lattice has very recently been obtained using the Bethe ansatz 
technique.\cite{solutionhofst} The fact that an 
infinite 2D problem 
(Eq. \ref{discrSchr}) can be reduced to an effectively 
1D problem (Eq. \ref{harpereq}) is instrumental in using the 
Bethe ansatz technique. 
 In a finite 2D lattice with open boundary
 conditions, a general solution is formally written as
\begin{equation}
\label{formalsoln}
\psi(x,y)=A(k_{y}) \ \psi(k_{y},x) \ \exp^{\imath k_{y} y} + A(-k_{y}) \ \psi(-k_{y},x) \ \exp^{-\imath k_{y} y},
\end{equation}
where $\psi(k_{y},x)$ and $\psi(-k_{y},x)$ are the formal solutions of 
Eq. \ref{harpereq}. 
The motion in the $x$ and $y$ spatial directions generally can not be separated in Eq.\ (\ref{formalsoln}). 
The difficulty encountered in obtaining an analytic solution for 
the spectrum of a finite lattice 
stems from our inability to factor out a $\psi(x)$ 
factor in Eq. \ref{formalsoln} corresponding to the opposite $k_{y}$ 
momenta of an infinite lattice. The origin of this is the 
 breaking of the time-reversal symmetry by the magnetic field. 
Eq. \ref{harpereq} is not invariant under 
the $k_{y} \rightarrow -k_{y}$ transformation. 
However, for the two limiting values of the magnetic 
flux $\phi=0$ and $\phi=0.5\ \phi_{0}$, the time-reversal symmetry is not broken, and the finite lattice problem can be analytically solved by 
reducing it  to two
effectively one-dimensional problems. 
The exact solution for these two limiting values of flux is given by
\begin{equation}
\label{limits}
\psi(x,y)=A(k_{y}) \ \psi(k_{y},x) \sin (k_{y}), 
\end{equation}
where $k_{y}$ takes on discrete values $k_{y}=\frac{\pi n}{L_{y}+1}$ ($n=1, 2, ..., L_{y}$) with $L_{y}$ being the extension of the sample in the
 $y-$direction. The $x$ part of the wavefunction in 
Eq. (\ref{limits}) satisfies the Harper equation,
\begin{equation}
\label{xlimits}
\psi(x+1) + \psi(x-1)+ 2\ 
\cos \left( 2 \pi \frac{\phi}{\phi_{0}} x\right) \cos (k_y) \psi(x)= - E/t\  \psi(x).
\end{equation}
It can easily be verified by  direct substitution that Eq. (\ref{limits}) 
is indeed a  solution of Eq. (\ref{discrSchr}) for the two limiting 
values of the flux,  $\phi=0$ and $\phi=0.5\ \phi_{0}$. 

At $\phi=0$, Eqs. (\ref{limits}) and (\ref{xlimits}) 
reduce to a trivially diagonalizable problem with a solution given by the superposition of two independent standing waves in the 
$x$ and $y$ directions, 
\begin{equation}
\label{zerofield}
\psi_{(k_{x},k_{y})}=A(k_{x}) A(k_{y}) \sin (k_{x}) \sin 
(k_{y}), {\displaystyle{and}} 
\end{equation}
\begin{equation}
\label{energyzero}
E_{(k_{x},k_{y})}= -2 t (\cos (k_{x}) +\cos (k_{y})).
\end{equation}
In Eqs. (\ref{zerofield}) and (\ref{energyzero}) 
the pseudomomenta $k_{x}$ and $k_{y}$ of a finite $L=L_{x} \times L_{y}$ lattice take the discrete values $k_{x(y)}=\frac{\pi n_{x(y)}}{L_{x(y)}+1}$, with the 
integers $n_{x(y)}=1,2, ..., L_{x(y)}$. The normalization 
constants are $A(k_{x(y)})=2/\sqrt{2L_{x(y)}+1} \ 
[1- \sin [(2 L_{x(y)}+1) k_{x(y)}]/$$[ (2 L_{x(y)}+1) \sin k_{x(y)}] ]^{-1/2}$. We refer to the quantum numbers $k_{x(y)}$ as the 
pseudomomenta because the translational invariance is broken in a finite lattice, and the momentum is not a good quantum number. 
At $\phi=0.5\ \phi_{0}$, 
the problem reduces to solving the algebraic equation 
$\Delta_{m}(E)=0$, where $\Delta_{m}$ for $m \ge 2$ obeys a recursion relation 
$\Delta_{m}=[E/t - 2 \cos (k_{y})] \Delta_{m-1} - \Delta_{m-2}$  with 
$\Delta_{1}=(E/t)^{2} - 4 \cos^{2} (k_{y})$ and 
$\Delta_{0}=E/t - 2 cos (k_{y})$. We note that at 
$\phi=0.5\ \phi_{0}$, Eq. (\ref{xlimits}) has the symmetries of a bipartite 
lattice, and therefore the eigenspectrum is highly 
degenerate at this value of the flux.

At $\phi =0$, we can characterize the spectrum using elementary group theory. A finite square $ L=\sqrt{L} \times \sqrt{L}$ lattice has the symmetries of the $C_{4v}$ spatial point group. It is invariant under the following symmetry operations: 
(a) a rotation of the whole lattice by $2 \pi$; 
(b) a rotation by $\pi$; (c) a rotation by $\pm \pi/2$; 
(d) a reflection around the vertical or horizontal  
axes of symmetry; and finally (e) a reflection about 
the two main diagonals of the lattice. These symmetry operations form $5$ classes, and therefore the eigenstates 
of Eq. (\ref{discrSchr}) belong to $5$ irreducible representations. The 
 degeneracies are immediately deduced from the character table\cite{grouptheory} of the group $C_{4v}$ given in Table 1. 
The spectrum is either
 singly-degenerate, belonging to the one-dimensional A1, A2, B1, or B2 representation, or doubly-degenerate belonging to the two-dimensional E representation. In the last row of Table 1 we 
show the characters of the reducible representation of the 
group $C_{4v}$ of a $3 \times 3$ lattice in the occupation basis.
 Using the character table of the irreducible representations and 
the characters of this reducible representation, we deduce that for
 the $3 \times 3$ system, 
group theory predicts 2 pairs of doubly-degenerate eigenvalues
 and 5 singly-degenerate eigenvalues. Similarly, the number of 
degenerate states is deduced for a $\sqrt{L} \times \sqrt{L}$ lattice, and the results are given in Table 2. An additional electron-hole symmetry in the 
$\sqrt{L} \times \sqrt{L}$ problem mixes the $\sqrt{L}$ singly-degenerate 
states with total pseudomomenta $k_{x}+k_{y}=\pi$ at zero energy. 
This can also be  
seen from Eq. (\ref{energyzero}). To summarize, for a finite square 
$\sqrt{L} \times \sqrt{L}$ lattice at $\phi=0$,
 we find $g=1$ and $g=2$ degenerate states in the spectrum
as well as $g=\sqrt{L}$ degenerate states in the middle of the spectrum (see Figures 2(a) and 4 at $\phi=0$ for the examples of these spectra). 
As $\phi \rightarrow 0$ the magnetic field effects can be calculated in perturbation theory, where the perturbing Hamiltonian is given by the total persistent current operator 
\begin{equation}
\label{perturbation}
H_{1}=- \imath t 2 \pi \frac{\phi}{\phi_{0}} 
\sum_{x=1}^{L_{x}} \sum_{y=1}^{L_{y}-1} x 
[{{c}^{\dag}}_{(x,y+1)} 
c_{(x,y)} - h.c.].
\end{equation}
At a small flux, the persistent current carried by the singly-degenerate states is linear in the flux in the lowest 
nonvanishing (2nd) order perturbation theory. The persistent current carried by the doubly-degenerate states is easily determined by a degenerate first order perturbation theory, where the 1st order correction to the energy is $E^{1}=\pm |\langle \gamma| H_{1}| \beta \rangle$ 
with $\gamma$ and $\beta$ being the two doubly-degenerate states with the same energy at $\phi=0$. 

We summarize our results for the limiting values of flux for the $3 \times 3$ system in Table 3. For generic values of the 
 flux, the spectrum of 
the $3 \times 3$ system is given by the solution of the eigenvalue problem of a $9 \times 9$ 
matrix. A symmetry operation which commutes with the 
finite two-dimensional square lattice Hamiltonian at $\phi \neq  0$ will reduce the problem. In a particular gauge such a symmetry operation should leave invariant the Peirels 
phase factors in Eq. (\ref{discrSchr}). In both the symmetric
 ($\vec{A}=\frac{B}{2}(y,-x,0))$ and the Landau 
($\vec{A}=B(0,x,0))$ gauge, upon transversing the distance between the two neighboring points $(x_{1},y_{1})$ and $(x_{2},y_{2})$ on a square lattice, the electron's wavefunction gains a phase proportional to $f(2,1)=y_{2}x_{2}-x_{1}y_{1}$. The object $f
(2,1)$ is invariant under $\sigma_{v}$ operations, which are
  reflections about the main diagonals of the lattice, i.e. 
$x \rightarrow y,\ y \rightarrow x$, and 
$x \rightarrow -y,\ y \rightarrow -x$. It is easy to verify that this symmetry is indeed observed in the numerically calculated current distributions in Fig. 5. In these gauges the $\sigma_{v}$ operations exhaust all the spatial symmetries, but we can not rigorously 
prove that this result is gauge invariant. We can, however,  comment on the degeneracies (or level crossings) which occur in the spectrum in Fig. 2(a) at $\phi=\frac{1}{8}\ n \phi_{0},\ n=1,\ 2,\ 3,\ 4$. The limiting cases are given in Table 3. The 
degeneracy at $\phi=0.25\ \phi_{0}$ at zero energy follows 
from the particle-hole symmetry. The values of flux at which 
the degeneracies occur, $\phi=\frac{1}{8}\ \phi_{0}$ and  $\phi=\frac{3}{8}\ \phi_{0}$, are the values when the first and second semiclassical orbits are commensurate with the lattice. 

In a one-dimensional ring of length $L$ enclosing a magnetic flux 
$\phi$, the symmetry operator which commutes with the Hamiltonian is the 
 magnetic translation operator along the ring, and the one-dimensional problem is easily diagonalized for all values of the flux with a spectrum given 
by\cite{cheung}
\begin{equation}
\label{ring}
E_{n}=- 2 t \cos 
\left[\frac{2 \pi}{L}\left( \frac{\phi}{\phi_{0}}+n\right)\right],
\end{equation}
where $n=1, ..., L$. At a finite flux, the eigenstates of the ring Hamiltonian 
 are also the eigenstates of the momentum, which are   
obtained from the momentum at $\phi=0$ by the same shift 
$\frac{2 \pi}{L}( \frac{\phi}{\phi_{0}})$ for all eigenstates. 
(As we shall show  later, this result also holds in an interacting 
one-dimensional Hubbard ring.) 
The energy level crossings between any two different $n$ and $n^\prime$ 
states occur at $\phi$ satisfying 
$n-n^{\prime}=2 \frac{\phi}{\phi_{0}}+ 2\ (integer)$. 
The resultant energy spectrum is periodic in 
$\phi$ (with a period $\phi_{0}$) through the ring. The 
discontinuities in the persistent current occur at $\phi=0$ ($\phi=0.5$) for even (odd) total number of electrons in the ring.\cite{cheung} A typical spectrum of a 1D tight-binding ring is shown in Fig. 7 for the
$L=9$ site ring.

It is of interest to compare the relative magnitudes 
of the 1D and 2D non-interacting persistent currents. In the 1D ring with N electrons the total current $I(N, \phi)$ is 
of the same order as the persistent current carried by the occupied individual states. This happens due to the cancelation of the persistent currents carried by the states with opposite momenta in the ground state distribution. 
In one period, $I(N, \phi)$ in the 1D ring is given by 
\begin{eqnarray}
\label{1dcurrent}
I(N_{odd})=-
I_{0}\sin \left(\frac{2\pi}{L}\frac{\phi}{\phi_{0}} \right) 
\frac{\sin(\pi N/L)}{\sin(\pi/L)}, \ -0.5 \le \frac{\phi}{\phi_{0}} \le 0.5,  
{\textstyle{and}} \\
I(N_{even})=
I_{0}\left[ \sin \frac{2\pi}{L}\left(N/2+\frac{\phi}{\phi_{0}}\right)
-\sin\left(\frac{2\pi}{L}\frac{\phi}{\phi_{0}}\right) 
\frac{\sin((N+1)\pi/L)}{\sin(\pi/L)}\right],\ 0 \le \frac{\phi}{\phi_{0}} \le 1,
\end{eqnarray}
where $I_{0}=e v_{F}/L$ with $v_{F}=2t/\hbar$. Similar 
formulas were derived in Ref.\onlinecite{cheung}.
In a 2D system, the magnitude of the total current can decrease
 due to the cancelation between opposite sign persistent current contributions 
coming from bulk and edge states. In our earlier work,\cite{kotlyarsdsPCUR}
we showed that the typical 
2D persistent current $\langle I^2 \rangle^{1/2}$ scales with the size of the boundary of the system. For completeness, we reproduce in Fig. 8 our results for the calculated system-size dependence of the
 typical current in the half-filled 1D and 2D systems with the same flux through the areas of each system. 
In the case of a constant flux density (i.e., the same flux piercing a unit cell for different systems), the magnetization density
(or the persistent current) saturates for large system sizes of 2D lattices for all electron fillings as shown in Fig. 9. 

\section{The interacting spectra}
\subsection{The minimal Hubbard model}
\subsubsection{Periodic $S_{z}$ oscillations}
A puzzling and interesting feature of the 
results plotted in Figs. 1 and 2 is  
the {\it{periodic}} oscillations of the $z$-component $S_{z}$ 
of the total spin in the ground state of a finite quantum dot array. 
The existence of spin flips in the ground state by itself
is not surprising. Any level crossing between states 
with different spins leads necessarily to a spin flip. The presence of another flux 
value within a magnetic period that leads to a reverse flip 
can not be assumed {\it{a priori}}. This implies that the 
lowest energy states with different $S_{z}$ can
 be degenerate for a range of the magnetic flux or 
at discrete flux values. 
We show in Fig. 10 the 
 magnetic flux dependence of the ground state energy of the 
$3 \times 3$ array in the minimal Hubbard model 
 (Eq. (\ref{interactions}))
 for values of N and M for which we find the $S_{z}$ oscillations in Fig. 2(b).
To obtain the results in Fig. 10, we set 
$\varepsilon_{i\uparrow}=\varepsilon_{i\downarrow}=0$, and 
 also leave out
the terms not contributing to the $S_{z}$ oscillations and  consider
an interaction of the local form
$V_{int}={\displaystyle{\sum_{i=1}^{L}}}
U \hat{\rho}_{i\uparrow}\hat{\rho}_{i\downarrow}$ in Eq. (\ref{interactions}).
The ground state energy with fixed N and M as a function of the flux 
appears to consist of segments of parabolas with their 
centers shifted along both the field and energy axes. 
For $N=2, \ 3,\ 4,\ 5,\ 6$ in Fig. 10, the energy 
parabolas belonging to different M values overlap for a range of flux, 
leading to periodic $S_{z}$ oscillations in Fig. 2(b) for a 
 non-zero Zeeman splitting. 
This result is not restricted to 2D systems. We find it to be 
 valid also in 1D rings.
We show in Fig. 11 the flux dependence of the lowest energy with different N and M 
for which we find $S_{z}$ oscillations in Fig. 2(c)
in the Hubbard ring with $L=9$ sites. 
In the rest of this subsection, we 
 solve the Bethe ansatz equations to study the $S_{z}$
 degeneracy of the 
 ground state of a 1D Hubbard ring and then generalize our results to two dimensions.

The energy and the canonical momentum of an L-site Hubbard ring with N electrons 
enclosing a magnetic flux $\phi$ are given by the Bethe ansatz solution
\begin{eqnarray}
\label{betheenergy}
E= -2 \ \sum_{j=1}^{N} \cos k_{j},\\
\label{bethemom}
P=\sum_{j=1}^{N} \left[ k_{j}-\frac{2 \pi}{L}\frac{\phi}{\phi_{0}} \right]=
\frac{2 \pi}{L} \left[
\sum_{j} I_{j} + \sum_{\alpha} J_{\alpha} \right].
\end{eqnarray}
(For the 
sake of brevity, we refer to the 
canonical momentum defined in Eq. (\ref{bethemom}) as the momentum 
from now on.) 
The energy and the momentum of the interacting system in 
Eqs. (\ref{betheenergy}) and (\ref{bethemom})
 are given by expressions similar to those for a non-interacting system, involving a summation over $N$ pseudomomenta $k_{j}$. The pseudomomenta $k_{j}$ describe the charge degrees of freedom that are coupled to the spin degrees of freedom with associated spin quantum 
numbers $\lambda_{\alpha}$ in an interacting system. 
The set of 
numbers $k_j$ and $\lambda_\alpha$, referred to as charge
 and spin rapidities, is found\cite{liebwu} by solving a set
 of coupled Bethe ansatz equations:
\begin{eqnarray}
\label{bethek}
L k_{j}=2 \pi I_{j} + 2 \pi \frac{\phi}{\phi_{0}} - \sum_{\beta=1}^{M} 
2 \tan^{-1} \left[ 
\frac{\sin k_{j}-\lambda_{\beta}}{U/4t} \right],\\
\label{bethelambda}
\sum_{j=1}^{N} 
2 \tan^{-1} \left[ 
\frac{\lambda_{\alpha} - \sin k_{j}}{U/4t} \right]=
2 \pi J_{\alpha}  + \sum_{\beta=1}^{M}
2 \tan^{-1} \left[ 
\frac{\lambda_{\alpha} -\lambda_{\beta}}{U/2t} \right].
\end{eqnarray}
The quantum numbers $I_{j}$ ($J_{\alpha}$) are integers 
if M is 
even 
(N-M is odd) and half-odd 
integers if M is odd (N-M is even), and $J_{\alpha}$ is 
restricted\cite{schulz} to a range $|J_{\alpha}| < (N-M+1)/2$. 
The ground state energy is obtained by taking the consecutive sets of
integers $I_{j}$ and $J_{\alpha}$, 
 and the lowest excitations are obtained 
by having holes in the ground state distribution of quantum numbers. 

The problem of the persistent current in a 1D Hubbard ring was studied earlier in Ref. \onlinecite{YuFowler}. These authors found that the system 
accommodates  magnetic flux by creating a 
magnon excitation with a hole in the ground state 
$\lambda_{\alpha}$ distribution. In Ref. \onlinecite{YuFowler} 
it was assumed that the spin 
excitations caused by the magnetic flux in the system remain spin waves 
for all values of $U/t$. This allowed one\cite{YuFowler} to conclude that 
for infinite $U/t$, the ground state energy has $N$ cusps in a 
magnetic period. 
For large but 
finite $U/t$ relevant to quantum dot arrays 
under investigation, the assumption that a magnetic field leads 
to a spin wave excitation in a 1D ring works  well.
In Appendix A, we reanalyze the ground state quantum number distribution 
of a Hubbard ring 
with particular values of  N and M that were singled out in 
Ref. \onlinecite{YuFowler}. In addition we analyze the ground 
state distributions of $M-1$ states to understand the origin of the 
periodic $S_{z}$ oscillations. 

To summarize our analysis of Appendix A, we find that a 1D Hubbard  system of $N$ interacting 
electrons  behaves as a {\em single particle} on a ring in a magnetic 
flux: the ground state corresponds to a sequence of states 
with the consecutive  values of the total momentum which are 
 defined by Eq. (\ref{bethemom}).  
The 
resulting ground state energy has $N$ intersecting parabolic segments per
flux period for large $U/t$ in  
agreement with Ref.\onlinecite{YuFowler}. The interacting system 
changes its total momentum as a function of the flux by creating a 
magnon excitation. We find from the numerical solution 
of the Bethe ansatz Eqs. (\ref{betheenergy}) - (\ref{bethelambda})
that the dynamics of this excitation (i.e. its location 
in the spectrum of spin rapidities) is determined by the dynamics of the total momentum that changes sequentially in multiples of $2 \pi / L$ from its minimum to its maximum value within one magnetic period. This last finding was not 
emphasized in the earlier analysis\cite{YuFowler} 
of the persistent current in a 
Hubbard ring.  This result follows
from Eqs. (\ref{betheenergy}) - (\ref{bethelambda}) in the
$U/N \rightarrow \infty$ limit. In the limit of the infinite interaction, the
pseudomomenta are given by
\begin{equation}
\label{uinfty}
k_{j}=\frac{2\pi}{L} \left[ \sum_{j^{\prime}}(1-\delta_{j j^{\prime}})
I_{j^{\prime}} + \frac{(L/2\pi)P}{N}+\frac{\phi}{\phi_{0}}\right].
\end{equation}
A sequential set of the total momentum states  which minimizes  charge
rapidities in Eq. (\ref{uinfty}) and, therefore, the ground state energy can
be chosen by considering both, either spin or charge, excitations
in the ground state
quantum number distribution. The energy cost to create a charge excitation
in the infinite interaction
limit in a 1D Hubbard ring is associated with the motion of
noninteracting spinless fermions and its magnitude is of the order of
 $t$. The energy cost to create a spin excitation is
determined by $t^{2}/U$, leading to spin excitations
being the most energy-efficient  way in the 1D Hubbard ring
to accommodate the enclosed magnetic flux.

The total spin $S$ may not change in a new total momentum ground state. Therefore, we find that all cusps  in the ground state energy within a magnetic period are 
 associated with  changes in the orbital quantum numbers, but
 not necessarily with changes of the total spin. We find the ground states with nonzero total spin $S$ to be the generic situation with respect to the magnitude of the interaction $U/t$, the electron filling, and the flux. Such ground states 
maintain their ($2S+1$) degeneracy with respect to the different values of $S_{z}$ over the range of the magnetic flux. This extended degeneracy is maintained until the  
total spin {\it{and}} momentum change in either of the different
$S_{z}$ ground states. 
In the presence of the extended total spin 
degeneracies  
a small nonzero Zeeman term ($\omega_{z}$) in the Hamiltonian in Eq. (\ref{allham}) leads  
to the periodic oscillations of the $z$-component $S_{z}$
of the total spin $S$ 
as shown in Figs. 1 and 2. The reason is the following. 
For small $\omega_{z} \neq 0$ the ground state is the
 $M=N/2+1$ state, whereas for 
$\omega_{z}=0$ the $M=N/2$ and $M=N/2+1$ states are
 degenerate. Beyond this region, the $M=N/2$ state is the 
lowest energy state, and the system changes its $S_{z}$ until 
it completes a period and the ground state again becomes $M=N/2+1$ state. 
In the 2D $3 \times 3$ array, similar extended total 
spin degeneracies with respect to the
 $M,\ M-1,\ M-2$ states lead to the $S_{z}$ oscillations in Figs. 1 (a), (b) and 2(b). 

\subsubsection{Spin polarization transition}

The critical magnetic field $B_{c}$ characterizing the spin polarization transition in the minimal Hubbard model is determined to be 
$g\mu_{B}B_{c} \approx 7.5 \frac{t^{2}}{U}$ and $\approx 11 \frac{t^{2}}{U}$ at half-filling 
in the $L=9$ ring and the $3 \times 3$ array, respectively 
(see insets to Figure 1). In an infinite system, the critical field $B_{c}$ corresponds to a field driven ferromagnetic transition and was calculated by several authors for a 1D Hubbard ring neglecting the
 orbital contribution\cite{carmelo}. For large $U/t$ (strong interaction)
  the critical magnetic field in the ring depends on the filling
 $n=N/L$ through the relation
 $g\mu_{B}B_{c}=
\frac{8t^{2}}{U} n [1-\sin(2\pi n)/(2 \pi n)]$. In the
 thermodynamic limit, the critical field at half-filling is 
$g\mu_{B}B_{c}=\frac{8t^{2}}{U}$. 
 Qualitatively, the high value ($\approx 11 \frac{t^{2}}{U}$) of the 
critical field in the finite 2D system is not surprising, since 
  we  expect the quantum 
fluctuations to have a larger disordering effect in 2D 
systems, raising the critical field.
More quantitatively, this result can be understood as follows. The critical field
 determines the gap for the triplet excitations in the 
ferromagnetic phase. At half-filling, the critical field is 
then determined by 
$g\mu_{B}B_{c}=E_{0}(N=L,M=N-1)-
E_{0}(N=L,M=N)$
 with $E_{0}$ being the lowest eigenvalue of Eq. (\ref{allham}) for 
 particular values of $N$ and $M$. The energy of the 
spin polarized state is trivial, 
$E_{0}(N,M=N)=-N g \mu_{B} B_{c}/2$. 
In a strongly interacting $M=N-1$ case, a spin down electron will `attract' 
a spin up hole. The energetically favorable configuration on a lattice 
corresponds to a single site being occupied by the spin down electron and the spin up hole, and all the other sites being
 singly occupied by the ferromagnetically aligned spin up electrons.
All the electrons contribute the trivial Zeeman term to the total 
energy, but the site with the spin down electron and its spin up neighbors 
also contribute the exchange energy $E_{exch}$. The total energy 
of the $M=N-1$ spin up and one spin down electrons is 
$E_{0}(N=L, M=N-1)=-(N-1) g \mu_{B} B_{c}/2+g \mu_{B} B_{c}/2+E_{exch}$. Therefore, the critical field is determined by the exchange 
energy, $g \mu_{B} B_{c}=E_{exch}$. The exchange energy is given by $E_{exch}=\frac{4t^{2}}{U} \ f(d)$ with $f(d)$ being the
number of spin up neighbors of the spin down electron. In a 1D ring 
 $f(d)=2$, and in large 2D systems on the average $f(d)=4$ in the nearest neighbor tight-binding square lattice. 
Note that our numerical result for $B_{c}$ in finite 1D rings 
($B_{c} \approx 7.5 t^{2}/U$) is reasonably close to the thermodynamic 
result ($B_{c} = 8 t^{2}/U$). 

\subsubsection{Metal-insulator transition and spin ordering}
The charge stiffness, $D_{c}$, which characterizes  
[$D_{c}=0$ ($\neq 0$) characterizes the Mott insulator (metal)]
the degree of localization in a Mott-Hubbard system is defined for a 
1D ring of L sites to be\cite{kohn,staffordmillis,scalapino,gebhard}
\begin{equation}
\label{dc}
D_{c}=\frac{L}{2}
 \frac{\partial^{2}E_{0}(\phi)}{\partial \phi^{2}}.
\end{equation}
Note that, by definition, the charge stiffness is the negative 
flux derivative of the persistent current: 
$D_{c} \sim \frac{\partial I}{\partial \phi}$. 
In Fig. 13 we compare the noninteracting and interacting
 persistent currents calculated in the minimal
 Hubbard model for the $3 \times 3$ array. The interacting 
persistent current is suppressed by orders of magnitude at half-filling,
 $n=N/(2L)=0.5$. This is a finite size manifestation of the
 Mott-Hubbard metal-insulator transition as $n \rightarrow
 0.5$ and $U/t \ \neq 0$. For small values of the flux the localization
 effects due
 to the magnetic field are small, and the magnetic response
 is not strongly affected by interaction at low
 filling in the metallic phase as can be seen in 
Figs. 13 (a)-(d). The additional discontinuities
 of the persistent current in Fig. 13 arise from the 
 total spin $S$
 or from accidental degeneracies that occur with no Zeeman term
 in the Hamiltonian. In the insulating phase  ($N=9$ in Fig. 13)
 the persistent current is suppressed by orders of magnitude from its noninteracting value. Apparently, the
 ground state is described by one set of quantum numbers in a magnetic period, leading to the absence of  discontinuities of the persistent current. The magnetic response is paramagnetic at zero flux. We verify this suppression of the magnitude and
 oscillations as well as the paramagnetic nature of the persistent current at half-filling  by doing calculations for $2\times 2$, 
$3 \times 2$, $4 \times 2$ and $3\times 3$ quantum dot arrays; these results are shown in Fig. 14. The rate of the
 suppression of the persistent current with the increased
 strength of the interaction $U/t$ in the half-filled 
$3 \times3$ array depends on the orbital magnetic flux. We
 compare this rate in the $3 \times 3$ array for the two values of
 the flux, $\phi/\phi_{0}=0.1$ and $\phi/\phi_{0}=0.4$ in Fig. 15. The magnetic field localization at $\phi/\phi_{0}=0.4$ enhances the effect of the Hubbard $U/t$ on the magnitude of the persistent current, compared to a slower decay of the persistent 
current at $\phi/\phi_{0}=0.1$. 

Our use of the terms `insulating' and `metallic' to 
describe the $N=L$ and $N<L$ finite Hubbard
 systems is justified on the basis of the existing 
work\cite{kohn,staffordmillis,scalapino,gebhard,assaad}. Using the linear response theory it can be shown that the charge stiffness at zero flux in a 1D ring is proportional to the zero-
temperature DC conductivity of the system, with the real part of the conductivity given
 by\cite{staffordmillis,scalapino}
\begin{equation}
\label{conductivity}
\sigma(\omega)=2 \pi D_{c}(\phi=0) 
 \delta(\omega)+\sigma_{reg}(\omega).
\end{equation}
The $\sigma_{reg}(\omega)$ term gives the weight of the finite
 frequency part of the conductivity spectrum, and the coefficient $D=2 \pi D_{c}=\pi e^{2} (n/m)$, called the Drude weight, 
gives the zero frequency weight of the 
conductivity spectrum. In infinite systems, an insulator is 
characterized by $D=D_{c}=0$, and a metal has
 $D>0$\cite{kohn}. However, in finite systems in the insulating 
phase $D_{c}$ is small but nonzero, and 
is expected to reduce exponentially 
with  increasing system size.
Stafford and Millis derived an explicit asymptotic form for the scaling 
function of the charge stiffness in a half-filled Hubbard 
$L$-site ring (with $L$ even) \cite{staffordmillis},
\begin{equation}
\label{dcscaling}
D_{c}(L,\phi=0)|_{ L \rightarrow \infty }=(-1)^{L/2+1}L^{1/2}
\tilde{D}(U/t) 
\exp[-L/\xi(U/t)].
\end{equation} 
In Eq. (\ref{dcscaling}) $\tilde{D}(U/t)$ is a U/t dependent number, and $\xi(U/t)$ is 
the localization length, that at small $U/t$ behaves as\cite{staffordmillis}
\begin{equation}
\label{loclength}
\lim_{U \rightarrow 0} \xi = \frac{2t+U/2\pi+ ...}{\Delta(U,t)}.
\end{equation}
In Eq. (\ref{loclength}) $\Delta(U,t)$ is the Lieb-Wu Hubbard
 gap\cite{liebwu} for the charge excitations in the system, 
that is determined by 
$\Delta_{c}(U,t)=E_{0}(N+1,M)+E_{0}(N-1,M)-2E_{0}(N,M)$. 
Stafford and Millis also 
showed\cite{staffordmillis} that in a
 1D Hubbard ring the equal time $T=0$ Green's function
 decays
 exponentially as $G(x) \sim exp(-|x|/\xi)$. Therefore, the
 decay of the persistent current and that of the single-particle
 Green's function are shown to be connected in a 1D
 Hubbard ring. In the 2D Hubbard model at half-filling,
 where  exact
 thermodynamic results are not available,  recent 
numerical work using Monte-Carlo methods\cite{assaad}
 demonstrated that the single-particle Green's function scales 
as $G(x) \sim exp(-|x|/\xi_{2})$ with a correlation length
$\xi_{2}$ showing a critical behavior different from the 
1D case. 
%(For a Hubbard system in contact with a
 %particles reservoir with a chemical potential $\mu$ within 
%the Hubbard gap $\Delta_{c}$, $\xi$ scales as 
%$\xi \sim |\mu - \Delta_{c}|^{-\nu}$ with
 %$\nu=\frac{1}{4}$ ($\frac{1}{2}$) and $\Delta_{c}\approx 0.67$ 
%($=0$) 
%in 2D (1D), for more details see Ref.\cite{assaad}.) 
For our purpose the important point is that a 2D Hubbard system at half-filling is a 
Mott insulator, so following the exact 1D result of Stafford and Millis 
we make the ansatz that the charge stiffness in a 2D finite system 
{\em also scales} in the same way as the 2D Green's function, 
$D_{c} \sim \exp{-l/\xi_{2}(U/t)}$ with $l\sim \sqrt{L}$ being the 
characteristic linear size of the system. This 
provides some rationale for our assertion 
that $N=L$ results for the finite clusters in 
Figs. 14 and 15 correspond to the insulating phase. 

In the $3 \times 3$ array, the persistent current flows along the 
perimeter of the cluster for all $N$, but the sign of the 
persistent current in the system can be either diamagnetic or
 paramagnetic. Therefore, a steady
paramagnetism that we find in the half-filled small arrays can 
not be understood using a simple analogy with the paramagnetism of the 
edge states in the corresponding noninteracting 
system\cite{peierlssurpr}. The sign of 
the noninteracting persistent current is paramagnetic in the 
half-filled $2 \times 2$, $4 \times 2$, and $3 \times 3$ arrays 
and is diamagnetic in the $3 \times 2$ array.
The negative charge stiffness 
in the interacting $2 \times 2$ array is consistent with the 
prediction of 
Eq. (\ref{dcscaling}) for a $L=4 \times \ integer$
 ring\cite{staffordmillis,staffordshastry}. We do not know whether this
 orbital 
paramagnetism is a generic feature of the finite 
 half-filled 
2D Hubbard lattices at zero temperature, but our finding is consistent 
with previous exact diagonalization studies of the optical
conductivity of  $4 \times 4$ periodic 2D Hubbard 
lattices which also found a negative Drude weight at 
half-filling\cite{dagotto}. 
 We have calculated
 the distribution of the local moments 
($\langle 0| \hat{S}_{z,i} | 0 \rangle$) in the half-filled
 finite lattices with the Zeeman term set equal to zero in the Hamiltonian. 
The ground state of the 
$3 \times 3$ array belongs to a doublet with the total 
spin $S=\frac{1}{2}$ and $S_{z}=\pm \frac{1}{2}$. At $U=0$, the 
uncompensated moment in the system is 
distributed along the diagonals of the lattice. At  finite 
$U$, the off diagonal dots gain 
opposite magnetic moments, and the system is 
antiferromagnetically ordered. We verified that the 
magnitudes of the neighboring moments show a tendency to equalize 
with  increasing $U/t$. Therefore, a many-body state 
in the $3 \times 3$ array at half-filling represents a finite-size 
realization of the Mott-Heisenberg insulator\cite{gebhard} 
(antiferromagnetic at a finite $U/t$) ground 
state in the system, 
with a Mott-Hubbard gap opening in the charge excitations 
(as manifested in the strong suppression of the 
persistent current in Figs. 13-15).
The ground state for the even $L$-lattices
is a singlet 
($S=0,\ S_{z}=0$), with all the local moments 
being very small. In the 1D $L=9$-site ring 
($S=\frac{1}{2},\ S_{z}=\pm \frac{1}{2}$), we do not find any 
ordering of the local moments. 
We conclude that
 in  small clean $2D$ systems the onsite
 interactions strongly suppress 
(possibly exponentially at $N=L$) the magnetic 
response as $N \rightarrow L$, and for $N < L$ 
the onsite interactions have a much weaker  
destructive effect.  

\subsection{The extended Hubbard model}
We consider first  onsite and nearest 
neighbor off-site interactions in Eq. (\ref{interactions}). We set 
$V_{ii}=U$, and $V_{ij}=V$ if $(ij)$ are nearest neighbor dots 
($V_{ij}=0$ otherwise). Eq. (\ref{allham}) then has the form of an
extended Hubbard model\cite{hirsch,wang,chat}.

The Hubbard $U$ by itself leads to a spin density wave 
(SDW) commensurate with the lattice periodicity at 
half-filling in a 2D lattice. This SDW state has uniform onsite charge 
density, $\langle \hat{\rho}_{i} \rangle =1$. It is easy to see 
from Eq. (\ref{allham}) that at 
half filling the off-site interaction $V$ prefers double occupancy on the alternating sites, e.g. 
$\langle \hat{\rho}_{i\ odd} \rangle =2$, and 
$\langle \hat{\rho}_{i\ even} \rangle =0$. The minimum 
energy configuration for a large nonzero $V$ is the charge 
density wave (CDW) state commensurate with the lattice. The competition 
between $U$ and $V$ leads to a SDW-CDW transition in both 1D and 2D 
half-filled 
systems\cite{hirsch,wang,chat} as $V$ is increased. 
The mean field phase 
diagram  is easily obtained by considering the strong coupling limits of  
$U$ and $V$ in Eq. (\ref{allham}). The energy of the SDW state 
in the half-filled system is 
$E_{SDW}=\frac{U}{2} N + \frac{V}{2} N_{nn}$ with 
$N_{nn}$ being the number of all pairs of nearest neighbors on a lattice. 
The corresponding energy of 
the CDW state is
 $E_{CDW}= 4 \frac{U}{2} \frac{N}{2}$. The phase
 transition line at half-filling in the extended Hubbard model 
 is consequently given by
\begin{equation}
\label{transline}
V=V_{c}=U \frac{N}{N_{nn}},
\end{equation}
with $V < V_{c}$ being the SDW state.
In the 2D lattices with open boundary conditions,
 $N_{nn}=2L-L_{x}-L_{y}$ yields for large $L$,  $V_{c}\approx \frac{U}{2}$, 
and in the 1D rings $N_{nn}=L$ with $V_{c}=U$. In the usual
 Hubbard model with periodic boundary conditions,
 the corresponding equations for the SDW-CDW phase transition are 
$V_{c}=\frac{U}{4}$ (2D) and $V_{c}=\frac{U}{2}$
 (1D). In the half-filled 1D  finite
 rings the charge stiffness maximizes at the
 transition line between SDW and CDW states\cite{wang}.
Intuitively, this seems plausible because the persistent current is  
 sensitive to quantum 
fluctuations, which are maximized at the phase transition.
 We find a similar maximization at half filling ($N=9$) 
of the magnitude of the
 persistent current in the $3 \times 3$ 2D array as
 a 
function of the off-site interaction $V/t$ (Fig. 16). The persistent 
current is strongly suppressed on both sides of the SDW-CDW 
transition while being enhanced at the transition point. The current
  also changes sign at the transition and is more strongly 
suppressed deep in the CDW phase for large $U/t$. The 
critical interaction strength $V_{c}=0.65$  at which the finite 
size SDW-CDW `transition' occurs in the 2D array does not 
depend on the orbital magnetic flux (for $U/t=10$), 
which is  qualitatively 
consistent with 
the mean field and strong coupling results. The shape of the 
peak in Fig. 16, however, depends strongly on the orbital magnetic flux -- 
the 
transition at a flux of $\phi/\phi_{0}=0.4$ showing a 
larger enhancement  than that at 
 $\phi/\phi_{0}=0.1$.

The SDW-CDW transition in a 1D extended Hubbard model 
is found\cite{hirsch} to be 
a second  order phase transition for 
$U/t < U_{c}/t =3$, and a first order transition for $U/t > U_{c}/t$.
For example, the order parameter of the CDW state 
$p=(1/N) {\displaystyle{\sum_{i}}} (-1)^{i} \langle \hat{\rho}_{i} \rangle$ 
is nonzero in 
both SDW and CDW phases for $U>U_{c}$\cite{hirsch,wang}. A recent mean
 field calculation of the SDW-CDW phase diagram finds the 
transition to be first order (i.e. discontinuous in the order parameter) for 
all values of $U$ and $V$\cite{chat}. In the $3 \times 3$ system 
finite size effects dominate for weak coupling, and the 
persistent current peak position depends on the orbital 
flux for $U/t \lesssim 3$. We have calculated the persistent current as a 
function of $V/t$ for 
$U/t=3,\ 5,\ 7,\ 10,\ 20,$ and $30$ in the half filled 
$3 \times 3$ array and obtained  
 the finite size SDW-CDW phase diagram, which is shown in Fig. 17.
 The slope of $\approx 0.68$ of the linear fit 
to the data in Fig. 17 approximately agrees with the
 value $3/4$ predicted within the strong coupling 
theory. 

With the inclusion of the longer range (beyond nearest neighbors) 
off-site interactions in our calculations,  
we find that the spin antiferromagnetic order is destroyed in the 
finite $3 \times 3$  array with  the electron 
density distribution remaining uniform.  
We set the values of $V_{ij}$ 
using the classical capacitance matrix 
formalism\cite{bookset,middleton} as explained in Section II, 
and show in Fig. 18 our calculated persistent current
 at half filling ($N=9$) as a function of $V/t= (U/t) (C/C_{g})$, 
with $U/t =(e^{2}/C_{g})/t=10$ being fixed in the $3 \times 3$ array. 
An important conclusion from Fig. 18 is that strong long range 
interactions could enhance the magnitude of the persistent current back 
to its non-interacting half filled 
value, thereby effectively negating the Mott-Hubbard 
localization effect. 
Whether this is a general result (i.e. valid even in the thermodynamic limit) 
or purely a finite size effect is not clear at this stage.
 The intermediate values of the 
long-range interaction do not significantly modify the persistent current for $N<L$.  
This can be seen in Fig. 19 where we compare the 
($U/t=10,\ V=0$) and ($U/t=10,\ V/t=5$) persistent current 
for various filling fractions in the $3 \times 3$ array. 
Our results can be understood on the basis of the dependence 
of the inverse capacitance matrix elements on the relative 
ratio of $C/C_{g}$. For simplicity, let us discuss this dependence 
in  a double-dot array. In the limit $C/C_{g} \ll 1$, the interdot 
and intradot interaction matrix elements are 
$V_{12}=U(C/C_{g})\rightarrow 0$ 
and $V_{11}=U$. For $C/C_{g} \gg  1$, these matrix elements are 
$V_{12}=U/2$ and $V_{11} = (U/2) (1+C_{g}/C) \rightarrow U/2$. The large ratio of 
$C/C_{g}$ in the latter case effectively describes  a single composite system. In this limit all elements of the inverse capacitance matrix become equal. Thus the electrons become totally uncorrelated, and the only effect of interactions is to give an overall charging energy to the system.
Our results suggest
that in a finite system with weakly screened interactions the SDW-CDW transition 
may not occur, but the persistent current could still be enhanced to its 
noninteracting value.  

\section{Disordered quantum dot arrays}

\subsection{Non-interacting case}
In this section we consider the effect of  random disorder  
on the persistent current in the finite 2D quantum dot lattices 
in the spin polarized regime (see Figs. 2(a), 4 and 7 as examples 
of clean system spectra in this regime). We include disorder 
in our calculation through a spin-independent parameter 
$W$ that denotes the half-width of a uniform distribution of 
random on-site quantum dot energies centered around $\Delta$. The 
random on-site single-particle energies are set as
\begin{equation}  
\label{spdisorder}
\varepsilon_{i}=\Delta + \delta \varepsilon_{i},\ \
  \delta  \varepsilon_{i} \ \epsilon \ [-W/2,W/2].
\end{equation}  

The introduction of disorder could, in principle,  lead to 
Anderson localization in the system, with all electronic 
states being exponentially localized in the 
presence of strong disorder\cite{lee,belitz}. 
In general, disorder ($W$) and interaction ($U$, $V$) compete 
in determining the electronic properties of the resultant 
 ($W,\ U,\ V\ \neq 0$) 
 Mott-Hubbard-Anderson model.
 Disorder introduces two 
related length scales in the problem: a mean free path $l(W)$ and 
a localization length $\xi(W)$. (In the absence of any interaction, in two
 dimensions for a weak disorder 
the localization length $\xi(W)$ depends exponentially\cite{lee} 
on the mean free path as $\xi=l\ \exp [\pi k_{F} l/2]$ with $k_{F}$ 
being the Fermi wavevector whereas in one dimension $\xi \sim l$.) 
Scaling and perturbative 
arguments 
predict that for weak disorder (and in the absence of interaction) 
the conductance of a
2D lattice of size $(L)^{2}$ depends logarithmically on the 
mean free path whereas for strong disorder it falls off 
exponentially with the system size as $\sigma(L) \sim \exp(-L/\xi)$. 
We find two similar weak and strong disorder 
regimes in the behavior of the persistent 
current as a function of the disorder strength $W$, indicating a 
 connection between the persistent current and the DC conductance 
of the finite 2D system. In one
dimension Anderson localization occurs (i.e. $\sigma(L) \sim \exp(-L/\xi)$) 
 in the presence
of any finite disorder\cite{lee}, and the persistent 
current amplitude is exponentially suppressed\cite{cheung} 
for all $W$. In Fig. 20 we 
 show
the calculated log-log plot  at half-filling of the rms
current $\langle I^{2} \rangle^{1/2}$, averaged
over 100 disorder realizations for each value of $W$, as a function
of the disorder strength $W$ for various array sizes
($3 \times 3, \ 4 \times 4,\ 5 \times 5,\  6 \times 6$). In plotting these
results, we have factored out a scale factor,
$n_{c}/L=(L^{1/2}-1)^{2}/L$, so that the results for various system sizes
fall on top of each other, showing approximate scaling with
system size and disorder. (The scale factor $n_{c}=(L^{1/2}-1)^{2}$
is the number of unit cells or plaquettes in each square array of size $L$.) 
The two dashed straight lines in Fig. 20 give the
best fits to weak and strong disorder scaled currents, leading to the following
empirical results for the effect of disorder on the persistent current:
$\langle I^{2} \rangle^{\frac{1}{2}} \  \ = \
 \left(L/n_{c} \right) \ g(W)$,
with the scaling function $g(W)$ being given by
\begin{eqnarray}
\label{disscaling}
 g(W) \sim \left\{
\begin{array}{ll}
W^{-\gamma}, \ \gamma=(6.4 \pm 2.8) \times 10^{-2},\ \ \  W < 1.55\  \pi t \\
W^{-\beta}, \ \beta=1.84 \pm 0.49, \ \ \ \ \ \ W > 1.55 \  \pi t.
\end{array}
\right.
\end{eqnarray} 
We note that the empirical scaling defined by Eq. \ref{disscaling} (and 
shown in Fig. 20) is consistent with well-known noninteracting scaling 
localization result of there being logarithmic 2D localization at weak disorder.

\subsection{Interacting case}
Random disorder effects on the persistent current 
are subtle in the presence of  onsite interactions in 
the $3 \times 3$ array: 
they depend both on the filling in the array and the 
relative ratio $U/W$. Away from half-filling, the random 
 disorder suppresses the magnitude of the current and 
smoothens its discontinuities arising from
 level crossings of energies with different orbital 
quantum numbers. The $S_{z}$ oscillations and the 
corresponding 
discontinuities in the current in the metallic 
phase persist in the presence of intermediate 
( $W \stackrel{\displaystyle <}{\sim} U$)  disorder. 
 Hubbard interaction effects are not particularly 
important at high disorder, which is what is seen in Fig. 20 
(asterisks) 
for  the typical interacting current even at half filling.      
The intermediate disorder ($W<U$) produces 
an antilocalization effect at half-filling by enhancing 
the persistent current from its finite $U$-suppressed 
value as can be seen in Fig. 21 (a) and (b). 
We also observe that the disordered persistent current is 
 enhanced in the presence of the long range interaction from 
its $U$-suppressed value as can be seen in Fig. 21(c). 
Finally we note that the presence of disorder does not reverse the  
roles of the weak and strong orbital flux: the relative behaviors 
of persistent current at $\phi/\phi_{0}=0.1$  and 
$\phi/\phi_{0}=0.4$ in Fig. 21 can be compared with those for 
clean systems plotted in Figs. 15 and 18.
  
\section{Conclusions} 
We calculate the persistent current and electron addition spectrum in
 coherent two-dimensional semiconductor quantum
 dot 
arrays and one-dimensional quantum dot rings 
pierced by an external
 magnetic flux, using the exact
 diagonalization and the Bethe ansatz techniques within an
 extended Mott-Hubbard Hamiltonian.
We find that the magnetization density of a finite multidot array is periodic in the magnetic flux. In weak fields, we find flux 
{\it{periodic}} oscillations in the $S_{z}$ component of the total spin $S$.
We have included in our model the effects of both intradot/interdot Coulomb interactions and random disorder. We find that the persistent current 
is suppressed 
 in the antiferromagnetic Mott-insulating phase.
The finite-size realization of the spin density  wave - charge density wave ordering transition has been found to maximize the 2D array 
persistent current at half-filling at the critical value of the nearest neighbor interaction, a behavior qualitatively similar 
 to the charge stiffness of 1D rings\cite{wang}. 
We obtain the phase diagram for the SDW-CDW transition.
We demonstrate that the electrostatic long-range interdot interactions 
enhance the magnitude of the Mott-insulating phase persistent current to its non-interacting system value.
We find that the noninteracting persistent current as a function of the random disorder strength exhibits a behavior similar to that of the conductivity\cite{lee} of the system: it is strongly suppressed only at large disorder strengths. The
 Anderson-Mott transition\cite{belitz} has been found to have a
 subtle effect on the persistent current in the $3 \times 3$ array at half filling: the
intradot Coulomb interaction less than or comparable to
 the disorder strength $W$ ($U\stackrel{\displaystyle <}{\sim}W$) increases the disordered system persistent current.

We believe that the  2D persistent current physics in the 
$3 \times 3$ array discussed in this paper has already been 
indirectly observed  in 
 transport measurements in a $3 \times 3$ array of strongly coupled 
quantum dots fabricated by means of square grid gate structures on top of a
GaAs/AlGaAs heterostructure\cite{lenssen}. In low magnetic fields 
$(B<1\ T)$ at $T \approx 40\ mK$ the three conductance minima were 
measured in Ref. \onlinecite{lenssen} at B=0.18 T, 
B=0.48 T, and B=0.80 T, and superimposed on these minima small oscillations
periodic in B with period of 12 mT, 21 mT, and 24 mT were seen\cite{lenssen} 
in the 
measured magnetoconductance of the 
$3 \times 3$ array. The large 
conductance 
minima 
were interpreted to be due to the classical localized orbits in the array, 
with the 
cyclotron radius of each orbit being equal to $R_{c}=m^{*} v_{F}/eB$,  
with $R_{c}=330\ nm$, $125\ nm$, and $75\ nm$ in the three minima, 
respectively. The measured period of small oscillations was attributed 
to a reduction of a single classical orbit Aharonov-Bohm period 
due to the coupling between all possible 
classical orbits with a given $R_{c}$ in the lattice. 
We  suggest here an alternative explanation of this experiment. 
First, the observed large conductance minima correspond to the first 
three Landau orbits being commensurate with the size of 
a single dot. The semiclassical cyclotron radii, 
$R_{n}=\sqrt{(\hbar/eB)(2n+1)}$, are 60.5 nm (n=0), 64 nm (n=1), and 
64 nm (n=2) for the three conductance minima, respectively. Therefore, 
we estimate the radius of each dot as 63 nm 
(which is  reasonably consistent with the other dimensions of the array). 
These minima 
should be observed for temperatures for which $\hbar \omega_{c}\le 2kT$ 
yielding 
$T \approx 1.8\ K$, which is consistent with  $T=2\ K$ where 
the large conductance minima start to be affected by 
the temperature\cite{lenssen}. We agree with the experimentalists that 
the shorter period oscillations are due to the coupling of the electron 
orbits in the 
lattice. But we think that the contributing orbits  originate from the 
quantum mechanically required   
 gauge invariance in the experimental phase coherent system. These orbits 
arise due to electron hopping between the localized states of each dot. 
We estimate that 
the magnetic flux corresponding 
to one flux quantum $\phi_{0}$ piercing a unit cell of the lattice 
is $\approx \ 46$ mT (with a lattice 
constant of $a\approx 300$ nm). For the density of the 2DEG and the 
estimated size of the dot, the number of electrons in each dot is $\approx 5$. 
Considering one mobile electron at the 
Fermi energy in each dot, the experimental 
$3 \times 3$ array approximately 
corresponds to our $3 \times 3$ 
array at half filling (in the situation of the weakly 
screened long-range Coulomb interactions). The persistent current 
in our case is found to 
oscillate with a period of $\phi/\phi_{0}=0.25$ and 0.5 (see Figs. 13(h)
and 19(h)), that corresponds to $\Delta B=11.5$ mT and $\Delta B=23$ mT, 
which are consistent with the experimentally measured values. 

Finally, we emphasize three essential limitations of our work which may (generally) restrict its quantitative applicability to 
{\em realistic} quantum dot arrays (we do believe that our results 
describe well the qualitative aspects of coherent and collective 
physics in semiconductor quantum dot arrays): 
(1) we ignore completely the complicated 
(and interesting) details of single dot electronic structure, approximating 
the energy level in each dot by two spin-split energy levels described by the 
simple Fock-Darwin-Zeeman model; (2) we use the Mott-Hubbard model 
in treating interaction and correlation effects (both intradot and interdot) -- 
such a simple parameterization of Coulomb correlations may not apply quantitatively in real quantum dot arrays; (3) we have considered
only small (no more than 9 dots) finite arrays, being limited entirely by 
the exponentially growing Hilbert space size in our extended Hubbard model.
 While we see no particular hope of going beyond these approximations in the near future, we should emphasize that these limitations do not 
substantially restrict the qualitative applicability of our results to the 
{\em coherent collective} physics of quantum dot arrays. For example,
currently fabricated {\em coherent} quantum dot arrays typically 
contain only two to four quantum dots, and therefore our finite size 
calculations are, in fact, perfectly appropriate. Also, the basic physics 
of Coulomb blockade and quantum fluctuations are entirely captured 
in our extended Mott-Hubbard model, and the complications of single 
dot energetics (beyond that of two-level Fock-Darwin-Zeeman physics) 
can usually be distinguished from the collective 
physics of interest to us. 
Finally, some of our non-essential approximations, 
for example, restricting to zero temperature and using a 
constant interdot hopping amplitude $t$, can easily be relaxed 
in future calculations if such a need arises. We emphasize 
here that these non-essential approximations do not affect 
our qualitative conclusions in any significant manner. 
We believe that the extended finite Mott-Hubbard model as used by us 
is the appropriate minimal model which should form the basis of 
discussing the collective and coherent physics in mesoscopic 
semiconductor quantum dot arrays.
We feel that the direct observation of  equilibrium persistent current
and spin oscillations in coherent quantum dot square lattices and rings will shed light on the interplay among the single dot physics, coherence, disorder, long- and short-range Coulomb interaction effects on quantum phase transitions that we discuss in this paper.

This work was supported by the U.S.-O.N.R.

\appendix
\section{Ground state energy of a Hubbard ring enclosing a magnetic flux}
Here we analyze the ground state quantum number distribution of a Hubbard ring that we obtain by solving numerically the Bethe ansatz Eqs. (\ref{betheenergy}) - (\ref{bethelambda}) for all 
$N<L$, $M=N/2$ and $M=N/2 -1$.

\paragraph{$N=4n+2$, $M=2n+1$, and $M=2n$.}

For $N=4n+2$  and $M=2n+1$ the ground state distribution at a flux 
$\phi \ge 0$ is given 
by\cite{YuFowler}:
\begin{eqnarray}  
\label{aI}
\{I_{j}\}=-\frac{(N-1)}{2}, -\frac{(N-1)}{2}+1, ..., \frac{(N-1)}{2}, 
{\displaystyle{and}} \\
\label{aJ}
\{J_{\alpha}\}=-\frac{(M-1)}{2}-1, -\frac{(M-1)}{2}, ..., 
-\frac{(M-1)}{2}+p-1, -\frac{(M-1)}{2}+p+1,...,\frac{(M-1)}{2}.
\end{eqnarray}
In Eq. (\ref{aJ}) the position $p$ of a hole  in the distribution 
of the spin rapidities varies from $p=0$ to $p=M$ 
within one half of a magnetic period, and $p=0$ corresponds to the ground 
state distribution at $\phi=0$. The ground state distribution
for negative values of the flux is obtained by shifting all 
$J_{\alpha}$ to the right by one unit. The momentum $P$ of the 
state is given by the momentum $q$ of the spin wave excitation 
$P=q=-\frac{2\pi}{L} \ p$. Thus, the total momentum is zero at zero 
flux and then takes the values of all 
consecutive multiples of $-\frac{2\pi}{L}$ 
within one half of magnetic period, with the momentum being equal to  
$\frac{L}{2\pi} P= -M = -\frac{N}{2}$ at $\phi/\phi_{0}=0.5$. 
The distribution in Eq. (\ref{aJ}) is valid for $N<L$, for $N=L$ the 
distribution of both the charge and spin quantum numbers remains
symmetric at a nonzero flux.     
Since  
$\frac{L}{2\pi} P_{max}=-\frac{N}{2}$,  
in the limit of large $U/t$ the ground state energy has $N$ 
cusps at $\phi/\phi_{0}=\pm (p+1) \frac{1}{2N}$, 
and the persistent current is diamagnetic at 
$\phi/\phi_{0}=\pm p \frac{1}{2N}$. It was proved by 
Stafford and Millis\cite{staffordmillis} that the $N=4n+2$ ground state 
of the electron system under periodic boundary conditions is   
a spin singlet (total spin $S=0$) at $\phi=0$, and it 
can be either a spin triplet (total spin $S=1$) or a singlet at 
$\phi=0.5\ \phi_{0}$ (a flux value where the antiperiodic 
boundary conditions are realized).  
In the $L=9$-site Hubbard ring we find that at $\phi=0$, 
the ground state is a spin singlet, and at $\phi=0.5\ \phi_{0}$ 
it is a spin triplet with $S=1$. At $\phi=0.5\ \phi_{0}$, the ground 
state, therefore,  has a $(2S+1)$ spin 
degeneracy.

The energy of the $N=4n+2$  and $M=2n$ ($S_{z}=1$) state is higher 
than the ground state energy of the singlet state at $\phi=0$ (as an example, 
see Fig. 11(d) for $N=6$, $M=3$, and $M=2$). 
The energy of the $S_{z}=1$ state at $\phi \ge 0$ is minimized by choosing 
the ground state distribution as 
\begin{eqnarray}
\label{aIM1}
\{I_{j}\}=-\frac{N}{2}, -\frac{N}{2}+1, ..., \frac{N}{2}-1,
{\displaystyle{and}} \\
\label{aJM1}
\{J_{\alpha}\}=-\frac{(M-1)}{2}, ...,
\frac{(M-1)}{2}-p, -\frac{(M-1)}{2}-p+2,...,\frac{(M-1)}{2}+1.
\end{eqnarray}
In Eq. (\ref{aJM1}) the position $p$ of a hole  in the distribution
of the spin rapidities varies from $p=M$ to $p=0$
within one half of a magnetic period, and $p=M$ corresponds to a ground
state distribution at $\phi=0$. The charge degrees of freedom contain a
nonzero momentum $\frac{L}{2\pi}r=-\frac{N}{2}$ in the system, and the  
 spin rapidities are positioned in the Fermi sea to minimize the total 
momentum at $\phi=0$. The  vacant hole in this system is  
also within the Fermi bounds. The total momentum of the state described 
by Eqs. (\ref{aIM1}) and (\ref{aJM1}) is $\frac{L}{2\pi} P= -(\frac{N}{2}-p)$. 
Thus, at $\phi=0$ and $\phi/\phi_{0}=0.5$, the  minimum and maximum momenta
are $\frac{L}{2\pi} P= -1$ and $\frac{L}{2\pi} P= -\frac{N}{2}$, 
respectively. Within 
our numerical accuracy we do not find that the excitation can create 
a zero total momentum at $\phi=0$ in the range of the interaction 
 $U/t$ values from 0 to 200. Therefore, at $\phi=0$, 
there is a cusp corresponding to the crossing of equal and opposite minimum 
momenta, and the ground state energy has a total of $N-1$ cusps per 
magnetic period. The spin quantum numbers of the $N=L$ ground state remain 
symmetric in a magnetic period. 
The $S_{z}=0$ and the $S_{z}=1$ states have the same total 
momenta at the last segments 
of the energy curves that includes $\phi=0.5\ \phi_{0}$. 
This is a situation for the $N=6,\ M=3$ and 
$N=6,\ M=2$ states in the $L=9$-site ring at $\phi=0.5\ \phi_{0}$ 
in Fig. 11(d). 
The total spin of the $M=2$ system does not change in a magnetic 
period and is equal to 1. The degenerate $M=3$ and $M=2$ states
 belong to 
a spin triplet in the segment of the energy curve centered at 
$\phi=0.5\ \phi_{0}$. 
We conclude that the magnetic flux changes the total momentum of 
a state sequentially by one unit from one parabola segment 
to another (for smaller values of $U/t$, particular 
values of the total momenta are missing in the ground 
state\cite{YuFowler}). Each energy parabola segment has a  
fixed value of the total spin, which may or may not change with 
a change of the total momenta. The extended 
degenerate nonzero total spin regions are not specific to the
 ground state of the $N=4n+2$ systems. 
They occur whenever 
the magnetic field drives the systems with different $S_{z}$
 through a series of the consecutive total momentum states to the
 flux region where the ground state has a nonzero total spin $S$
 and the different $S_{z}$ states thus have the same total momenta.

\paragraph{$N=4n$, $M=2n$, and $M=2n-1$.}

For $N=4n$ ($N<L$) and $M=2n$ the ground state distribution at a flux 
$\phi \ge 0$ is the same as for $N=4n+2$ and  $M=2n$ case that is 
given in Eqs. (\ref{aIM1}) and (\ref{aJM1}).  
In the current situation, a spin wave excitation 
is above the Fermi sea, and the minimum momentum is 
$\frac{L}{2\pi} P= -(\frac{N}{2}-M)=0$ at $\phi=0$. The ground state 
energy has $N$ cusps in a period,\cite{YuFowler} and the 
persistent current is diamagnetic at the $\phi/\phi_{0}=\pm p \frac{1}{2N}$.
The ground state distribution for $N=4n$ and $M=2n-1$ is given in 
Eqs. (\ref{aI}) and (\ref{aJ}). For large $U/t$, an excitation 
above the negative Fermi sea of spin rapidities 
can be created, so the momentum at $\phi/\phi_{0}=0.5$ is 
$\frac{L}{2\pi} P= - \frac{N}{2}$.  
 The $S_{z}=0$ and $S_{z}=1$ states can have the same 
minimum momenta $P=0$ at $\phi=0$. In Fig. 11(c) for $N=4$ electrons 
in the ring, the energy parabolas of $M=2$ and $M=1$ belonging 
to the spin triplet ($S=1$) states are coincident 
in the $P=0$ magnetic flux region centered at $\phi=0$. For $N=8$ 
in Fig. 11(e) the levels of the $P=0$ momentum triplet state  
($S=1,\ S_{z}=1$) and 
$\frac{L}{2\pi}P=\pm 4$ singlet state ($S=0,\ S_{z}=0$) cross in the 
vicinity of $\phi=0$ and  
are non degenerate at $\phi=0$.

\paragraph{$N=4n+1$, $M=2n$, and $M=2n-1$.}

For $N=4n+1$ ($N<L$) and $M=2n$ electrons in a Hubbard ring, both sets of $I_{j}$ 
and $J_\alpha$ are integer numbers. To accommodate the variation of the 
total momenta, two holes are present in the ground state 
distribution at a nonzero positive flux: 
\begin{eqnarray}  
\label{cI}
\{I_{j}\}=-\frac{(N-1)}{2}, -\frac{(N-1)}{2}+1, ..., \frac{(N-1)}{2}, 
{\displaystyle{and}} \\
\label{cJ}
\{J_{\alpha}\}=-\frac{(M-2)}{2}, ...,
-\frac{M}{2}+p_{2}-2, -\frac{M}{2}+p_{2},...,0,..., 
p_{1}-1, p_{1}+1,...,\frac{M}{2}.
\end{eqnarray}

In Eq. (\ref{cJ}) the position $p_{1}$ of a hole  in a distribution 
of spin rapidities varies from $p_{1}=0$ at 
$\phi=0$ to $p_{1}=\frac{M}{2}$ with $p_{2}$ being fixed at $p_{2}=0$. 
While the hole $p_{1}$ transverses from $0$ to $\frac{M}{2}$, the 
momentum varies from $0$ to $-\frac{M}{2}$. A further increase  
of the momentum is accomplished through the motion of the second 
hole $p_{2}$ from $0$ to $\frac{M}{2}-1$ with $p_{1}$ 
being fixed at $\frac{M}{2}$. The total momentum  
becomes $P=-\frac{2\pi}{L}M$ 
at $\phi/\phi_{0}=0.5$. Therefore, in the ground state distribution 
there is always one hole which moves.  We show the energies for 
consecutive  values  of the total momentum in a 
$L=15$ ring for $n=2$ and $L=9$ ring for $n=1$ 
for $U/t=200$ in Fig. 12 (a) and 
(b), respectively. The ground state energy curves in Fig. 12 (a) and (b) consist of the $N$ consecutive momentum states within a magnetic period.
The earlier analysis 
of the persistent current in a Hubbard ring did not  
emphasize the dynamics of the total momentum, and 
Yu and Fowler 
mistakenly concluded that their Eq. (3.13) describes the ground 
state distribution for any nonzero flux, and that at the one 
half flux quantum all the $J_{\alpha}$'s have to be consecutive 
integers\cite{YuFowler}. 

The charge and spin quantum numbers are half-odd 
integers for  $N=4n+1$  and $M=2n-1$ electrons in a ring. 
The ground state distribution at $\phi \ge 0$ for $n>1$ is given by 
\begin{eqnarray}
\label{cIM1}
\{I_{j}\}=-\frac{N}{2}, -\frac{N}{2}+1, ..., \frac{N}{2}-1,
{\displaystyle{and}} \\
\label{cJM1}
\{J_{\alpha}\}=-\frac{M}{2}, ...,
-\frac{M}{2}+p_{2}-1, -\frac{M}{2}+p_{2}+1,...,-\frac{1}{2},...,
-\frac{3}{2}+p_{1}, \frac{1}{2}+p_{1},...,\frac{M}{2}+1.
\end{eqnarray}
In Eq. (\ref{cJM1}) the hole $p_{1}$ moves from 0 to $(M+3)/2$ with 
$p_{2}$ being fixed at $p_{2}=0$, and then the hole 
$p_{2}$ moves from $0$ to $(M-3)/2$ while $p_{1}=(M+3)/2$.
The motion of the holes in the distribution of the spin rapidities 
leads to a cancelation of the $-N/2$ momentum concentrated in the 
charge degrees of freedom at zero flux and the consequent 
integral decrease of the total momentum. In the $L=9$ ring 
with $N=5$ electrons, for the value of the interaction 
$U/t=10$, the states $M=2$ and $M=1$ have different values of 
the total spin $S=\frac{1}{2}$  and  $S=\frac{3}{2}$  and 
 the total momenta $\frac{L}{2\pi} P=-1$ and   
$\frac{L}{2\pi} P=-2$, respectively. These states are nondegenerate, 
and there are no oscillations of the total spin $S_{z}$ in the ground 
state of the $L=9,\ N=5$ Hubbard ring.

\paragraph{$N=4n-1$, $M=2n-1$, and $M=2n-2$.}
The lowest energy state for $N=4n-1$, $M=2n-1$ is obtained 
by choosing the distribution of quantum numbers that is given 
in Eqs. (\ref{cIM1})-(\ref{cJM1}). At zero flux, the hole 
$p_{1}$ starts moving from $J_{\alpha}=\frac{1}{2}$, and it 
moves $(M+1)/2 - 1$ consecutive steps to the right. In the remaining 
flux region the hole $p_{2}$ transverses to the right in 
$(M-1)/2 - 1$ consecutive steps. The distribution in the
$M=2n-2$ case is given in Eqs. (\ref{cI})-(\ref{cJ}). In this case the 
hole $p_{2}$ moves until the total momentum becomes 
$-(N+1)/2$ at $\phi/\phi_{0}=0.5$. At $\phi=0$ for $N=3$ 
electrons in a $L=9$-sites ring in Fig. 11(b), the two 
lowest $M=1$ and $M=0$ energy states with momentum $P=0$ and 
the total spin $S=\frac{3}{2}$ are degenerate 
in the $\phi=0$ region.

\begin{figure}
\caption{
(a) [(c)] A chemical potential $\mu_{N}$ as a 
function of flux $\phi$ in units of $\phi_{0}=h/e$ through an 
elementary cell [a ring in (c)] of a $3\times 3$ quantum dot 
lattice [$9-$site ring in (c)] 
with $N(=1-9)$ excess quasiparticles in the minimal
 Hubbard model approximation; (b) [(d)] A corresponding 
$z$-component $S_{z}$ of the ground state total spin as a 
function of $\phi$. The magnetic flux is 
rescaled by $1/32$ to show the entire dynamics of the spectrum on one 
flux scale as explained in the text. 
Insets: The critical magnetic flux $\phi_{c}$ 
of the spin polarization transition as a function of the electron filling 
$n=N/(2L)$ in the arrays.}
\label{1}
\end{figure}
\begin{figure}
\caption{
(a) A chemical potential spectrum plotted versus flux through a unit cell of a non-interacting $3\times 3$ array with $\varepsilon_{\downarrow}=\Delta$, $N=M$ in Eq. (1); (b) [(c)] A chemical potential as a function of flux (top panel), and the
 corresponding $S_{z}$ component of the
ground state total spin (bottom panel) of a $3\times 3$ array 
[$9-$site ring in (c)]
 with
 $\varepsilon_{\downarrow}=\Delta, \ 
\varepsilon_{\uparrow}=\Delta(1+\delta/\Delta)$ 
($\delta/\Delta=0.03$) in Eq. (1). (a), (b) and (c) can 
be directly compared with high and low field regimes in 
Figs. 1(a) and (c), respectively.}
\label{2}
\end{figure}

\begin{figure}
\caption{A spectrum of Eq. (16) plotted for rational fractions of flux $\phi/\phi_{0}=p/q$ ($q=120$, $p$ is incremented from 1 to 60) through a unit cell of an infinite tight-binding lattice. The first two Landau bands ($n=0, \ 1$) are marked on the plot.
}
\label{3}
\end{figure}
\begin{figure}
\caption{ A spectrum of a finite $15 \times 15$ tight-binding lattice versus $\phi/\phi_{0}$ through a unit cell. The different regions that are marked (a)-(l) on the plot are discussed in the text.}
\label{4}
\end{figure}
\begin{figure}
\caption{The persistent current and charge density distribution of the single-particle states marked in Fig. 4. 
The darkness of circles, and darkness, or thickness of connecting lines are 
proportional to the 
 magnitude of charge and current, respectively. Arrows on the 
lines show a direction of the persistent current. The maximum current and 
charge in (a) to (l) in units of $(et/h)$ and $e$, respectively, 
are 0.028, 0.009; 0.114, 0.018; 0.232, 0.023; 0.250, 0.102; 0.393, 0.039; 
 0.150, 0.030; 0.162, 0.062; 0.172, 0.019;  0.126, 0.011; 0.161, 0.027; 0.073,
0.015; 0.099, 0.017.}
\label{5}
\end{figure}
\begin{figure}
\label{6}
\caption{The total persistent current $I$ calculated using Eq. (15) for (a) $N=26$ electrons on a $15 \times 15$ lattice((e) in Figs. 4 and 5), and (b) $N=73$ ((h) in Figs. 4 and 5).}
\end{figure}
\begin{figure}
\label{7}
\caption{ The energy spectrum of a $9$-site tight-binding ring shown for one period of flux through the area 
of a ring; this spectrum can be compared with the high field regime in Fig. 1(c).}
\end{figure}
\begin{figure}
\label{8}
\caption{ $\langle I^{2} \rangle^{1/2}$ at half filling 
versus system sizes (L) ($L=L^{1/2} \times L^{1/2}$ with $L^{1/2}$ varying from 2 to 10): triangles (2D array), 
squares (1D ring). (The average is taken over the range of the same total flux through the systems, which varies from 
$-\phi_{0}/2$ to $\phi_{0}/2$.) The dashed lines are the best linear fits to the dependence $\langle I^{2} \rangle^{1/2} \sim L^{\alpha}$ with $\alpha=0.46\ (1.1)$ in a 2D (1D) systems.}
\end{figure}
 \begin{figure}
\label{9}
\caption{A typical current $\langle I^{2} \rangle^{1/2}$, $I$ at $\phi=0.25\ \phi_{0}$, and the maximum current 
$I_{max}$ are shown from top to bottom, respectively, as a function of $N$ electrons in $L^{1/2} \times L^{1/2}$ systems (with $L^{1/2}$ varying from 2 to 10) with the same flux density for systems of different sizes.} 
\end{figure}

\begin{figure}
\label{10}
\caption{The energy of the
 $3\times 3$ quantum dot array calculated using Eq. (1) in a minimal Hubbard model approximation ($\frac{V_{ii}}{2}=U=10t,\ V_{ij}=0,\ 
\varepsilon_{\uparrow}=\varepsilon_{\downarrow}=0$)
as a function of the flux through the unit cell. The energy
 is plotted for all values of $N$ and $M$ which are found
in the ground state of the array in Fig. 2(b). In (a)-(f), a
 number of electrons $N$ in the array is as indicated in the
 legend; $M=N/2$, $M=N/2 -1$, $M=N/2 -2$ energy states
 are shown with the solid, dotted, and dashed lines,
 respectively. The states with different $M$ that are degenerate 
over the range of flux have 
the total spin in (a) $S=1$; (b) $S=\frac{3}{2}$; (c) $S=2,\ 1,\ 0$; 
(d) $S=\frac{3}{2}$; (e) $S=1$.}
\end{figure}
\begin{figure}
\label{11}
\caption{The energy of an $9-$site Hubbard ring as a
 function of the flux through the ring calculated using the Bethe ansatz Eqs. (28)-(31). All parameters 
and plotting conventions are as given in the caption in 
Fig. 10. A sequence of states in one magnetic period, for the flux $-0.5 \le \phi/\phi_{0} \le 0.5$ is given by 
(a) $\frac{L}{2\pi}P=1, 0, -1$ ($M=1$) and 
$\frac{L}{2\pi}P=1, -1$ ($M=0$); 
(b) $\frac{L}{2\pi}P=1, 0, -1$ ($M=1$) and 
$\frac{L}{2\pi}P= 0$ ($M=0$); 
(c) $\frac{L}{2\pi}P=2, 0, -2$ ($M=2$) and 
$\frac{L}{2\pi}P=1, 0, -1$ ($M=1$); 
(d) $\frac{L}{2\pi}P=3, 0, -3$ ($M=3$), and 
$\frac{L}{2\pi}P= 3, 1, -1, -3$ ($M=2$); 
(e) $\frac{L}{2\pi}P=4, -4$ ($M=4$) and 
$\frac{L}{2\pi}P= 0$ ($M=3$).  The states with different $M$ 
that are degenerate
over the range of flux have
the total spin in (a) $S=1$; (b) $S=\frac{3}{2}$; (c) $S=1$;
(d) $S=1$.}
\end{figure} 
\begin{figure}
\label{12}
\caption{The energy of the 1D Hubbard rings versus the flux is shown for $N=4n+1,\ M=2n$ case for $L=15,\ N=9,\ M=4,\ U=200\ t$ (a) and $L=9,\ N=5,\ M=2,\ U=200\ t$ (b) for the consecutive values of the total momentum in the rings. Each energy curve is labeled in
 the plot by the value of its total 
momentum expressed in units of $\frac{L}{2\pi}$; in (b) it is also labeled by 
the value of the total spin $S$.}
\end{figure}
\begin{figure}
\label{13}
\caption{The persistent current in the $3 \times 3$ array of quantum dots as a function of flux through a unit cell for $U/t=0$ 
(solid line) and $U/t=10$ (dashed-dotted line) for $N=2-9$ in (a)-(h), respectively. The array is modeled in the minimal Hubbard approximation, with other parameters set the same as in Fig. 2(b). The corresponding $z$-component
 $S_{z}$ of the total spin in a ground state is shown in the bottom part of the plots, with an arbitrary offset for the illustrative purposes.}
\end{figure}
\begin{figure}
\label{14}
\caption{The persistent current at half-filling in the 
$2 \times 2$, $3 \times 2$, $4 \times 2$ clusters of quantum dots with $S=0,\ S_{z}=0$
 ground state and in the $3 \times 3$ lattice with
 $S=\frac{1}{2},\ S_{z}=\frac{1}{2}$ ground state as a 
function of flux through the unit cell. A key to each curve is indicated in the legend. All the parameters set the same as in Fig. 2(b).}
\end{figure}
\begin{figure}
\label{15}
\caption{The persistent current as a 
function of the Hubbard $U/t$ in the $3 \times 3$ array with $N=L=9$ 
at $\phi/\phi_{0}=0.1$ (solid) and $\phi/\phi_{0}=0.4$ (dotted). The same parameters used as in Fig. 2(b).}
\end{figure}
\begin{figure}
\label{16}
\caption{The persistent current as a function of the nearest neighbor 
off-site interaction $V/t$ in the 
$3 \times 3$ array with $N=L=9$ and $U/t=10$
at $\phi/\phi_{0}=0.1$ (solid) and $\phi/\phi_{0}=0.4$ (dotted). 
The same parameters used as in Fig. 2(b).}
\end{figure}
\begin{figure}
\label{17}
\caption{A phase diagram of the finite size SDW-CDW transition in the 
$3 \times 3$ array with $N=L=9$ is shown for the onsite and off-site nearest 
neighbor interaction $U/t$ and $V/t$ 
at $\phi/\phi_{0}=0.1$ (squares) and $\phi/\phi_{0}=0.4$ (triangles).
The same parameters used as in Fig. 2(b).}
\end{figure}
\begin{figure}
\label{18}
\caption{The persistent current as a function of the long range 
off-site interaction $V/t=(U/t) (C/C_{g})$ in the 
$3 \times 3$ array with $N=L=9$ and $U/t=10$
at $\phi/\phi_{0}=0.1$ (solid) and $\phi/\phi_{0}=0.4$ (dotted). The 
same parameters used as in Fig. 2(b).}
\end{figure}
\begin{figure}
\label{19}
\caption{The persistent current in the $3 \times 3$ array of quantum dots as a 
function of flux through a unit cell for $U/t=10$ and long range interaction 
$V/t=5$
(solid line) and $U/t=10,\ V/t=0$ (dashed-dotted line) for $N=2-9$ in (a)-(h), 
respectively.
 Other parameters are set the same as in Fig. 2(b). 
The corresponding $S_{z}$-component of the ground state
total spin is shown in the bottom part 
of the plots, with an arbitrary offset for the illustrative purposes.}
\end{figure}
 
\begin{figure}
\label{20}
\caption{
A log-log plot of $\langle I^{2} \rangle^{1/2}$
versus $W$
averaged over 100 disorder realizations
for each value of $W$ is shown
 ($U\ = \ V\ = \ 0$)
at half-filling in
 $3 \times 3$ (crosses), $4 \times 4$
(triangles), $5 \times 5$ (solid line), $6 \times 6$
(pluses).
Asterisks show the interacting ($U/t=10, V=0$) results
at half-filling for the $3 \times 3$ system.}
\end{figure}
\begin{figure}
\label{21}
\caption{The persistent current as a function of (a) and (b) 
the Hubbard $U/t$, (c) the long range
off-site interaction $V/t=(U/t) (C/C_{g})$ (with $U/t=10$)  in a
disordered $3 \times 3$ array with $N=L=9$ and 
at $\phi/\phi_{0}=0.1$ (solid) and $\phi/\phi_{0}=0.4$ (dotted). The 
persistent current is averaged over 10 disorder 
realizations. The disorder strength $W$ is indicated in the plots. The
same other parameters used as in Fig. 2(b).}
\end{figure}
\begin{table}
\begin{tabular}{|c|c|c|c|c|c|}
   & $E$ & $C_{2}$ & $2C_{4}$ & $2 \sigma_{v}$ & $2 \sigma_{d}$ \\ \hline
A1 & 1   &  1      &  1       &  1             &  1\\      
A2 & 1   &  1      &  1       &  -1            & -1\\
B1 & 1   &  1      &  -1      &  1             & -1\\
B2 & 1   &  1      &  -1      &  -1            & 1\\
E  & 2   &  -2     & 0        &  0             & 0\\  \hline
$\chi$ &9& 1       & 1        & 3              & 3\\ \hline 
\end{tabular}
\caption{Character table of $C_{4v}$ taken from Ref. [27]. The last row 
gives characters of $C_{4v}$ on a $3 \times 3$ lattice in the occupation 
basis.}
\end{table}
\begin{table}
\begin{tabular}{|c|c|c|}
   & $L_{odd}$ & $L_{even}$ \\ \hline
m(A1) & $(L+4 \sqrt{L}+3)/8$   & $(L+4 \sqrt{L})/8$   \\
m(A2) & $(L-4 \sqrt{L}+3)/8$   & $(L-4 \sqrt{L})/8$ \\
m(B1) & $(L-1)/8$       & $L/8$ \\
m(B2) & $(L-1)/8$       & $L/8$ \\
m(E)  & $(2L-2)/8$      & $2L/8$ \\  \hline
\end{tabular}
\caption{ A number $m(R)$ of eigenvalues belonging 
to an irreducible representation $R$ of the $C_{4v}$ group on a 
$L^{1/2} \times L^{1/2}$ lattice.} 
\end{table}
\begin{table}
\begin{tabular}{|c|c|c|c|c|}
\multicolumn{4}{|c|}{$\phi=0$}& 
\multicolumn{1}{c|}{$\phi=0.5 \ \phi_{0}$} \\ \hline \hline
$E/t$ & $(k_{x},k_{y})$ & Representation, &lowest order & $E/t,\ k_{y}$\\
      &                 & parity          &correction   &  \\ \hline
$-2\sqrt{2}$&$(\frac{\pi}{4},\frac{\pi}{4})$&A1, +&
$+\frac{\pi^2}{4}\left(\frac{\phi}{\phi_{0}}\right)^2$&$-2$, $\frac{\pi}{4}$\\
$-\sqrt{2}$ &$(\frac{\pi}{4},\frac{\pi}{2})$&E, -&
$-\pi \frac{\phi}{\phi_{0}}$&$-2$, $\frac{3\pi}{4}$\\
$-\sqrt{2}$ &$(\frac{\pi}{2},\frac{\pi}{4})$&E, -&
$+\pi \frac{\phi}{\phi_{0}}$&$-\sqrt{2}$, $\frac{\pi}{2}$\\
$0$        &$(\frac{\pi}{4},\frac{3\pi}{4})$&Mix &
$-2\pi \frac{\phi}{\phi_{0}}$&$-\sqrt{2}$, $\frac{3\pi}{4}$\\ 
$0$         &$(\frac{\pi}{2},\frac{\pi}{2})$&of  &
0                           &  $0$, $\frac{\pi}{2}$ \\
$0$        &$(\frac{3\pi}{4},\frac{\pi}{4})$&A1,B1,B2, +&
$+2\pi \frac{\phi}{\phi_{0}}$&$\sqrt{2}$, $\frac{\pi}{4}$\\
$\sqrt{2}$ &$(\frac{\pi}{2},\frac{3\pi}{4})$&E, -&
$-\pi \frac{\phi}{\phi_{0}}$&$\sqrt{2}$, $\frac{\pi}{2}$\\
$\sqrt{2}$ &$(\frac{3\pi}{4},\frac{\pi}{2})$&E, -&
$+\pi \frac{\phi}{\phi_{0}}$&$2$, $\frac{\pi}{4}$\\
$2\sqrt{2}$&$(\frac{3\pi}{4},\frac{3\pi}{4})$&A1, +&
$-\frac{\pi^2}{4}\left(\frac{\phi}{\phi_{0}}\right)^2$&
$2$, $\frac{3\pi}{4}$\\ \hline 
\end{tabular}
\caption{A summary of the analysis of the $3 \times 3$ 
tight-binding lattice spectrum for the limiting values of the flux.}
\end{table}
\end{document}